\documentclass[fleqn,usenatbib]{mnras}

\usepackage{fix-cm}
\usepackage{newtxtext,newtxmath}

\usepackage[T1]{fontenc}

\DeclareRobustCommand{\VAN}[3]{#2}
\let\VANthebibliography\thebibliography
\def\thebibliography{\DeclareRobustCommand{\VAN}[3]{##3}\VANthebibliography}

\usepackage{graphicx}	
\usepackage{amsmath}	
\usepackage{relsize} 
\usepackage{amssymb} 
\usepackage{bm}
\usepackage{physics}
\usepackage{mathtools}
\usepackage{xspace}
\usepackage{xcolor}
\usepackage{dsfont}

\newcommand{\HI}{\ion{H}{I}}

\newcommand{\CII}{\ion{C}{II}}
\renewcommand{\*}[1]{\boldsymbol{#1}}
\newcommand{\dinit}{\delta_{\scriptscriptstyle \mathrm{init}}}
\newcommand{\dXXI}{\delta_{\scriptscriptstyle \mathrm{21cm}}}
\newcommand{\dCO}{\delta_{\scriptscriptstyle \mathrm{CO}}}
\newcommand{\dtrue}{\delta_{\scriptscriptstyle \mathrm{init}}^{\scriptscriptstyle  \mathrm{true}}}
\newcommand{\dinitrec}{\delta_{\scriptscriptstyle \mathrm{init}}^{\scriptscriptstyle  \mathrm{rec}}}
\newcommand{\dinitCO}{\dinit^{\scriptscriptstyle \mathrm{CO}}}
\newcommand{\dinitXXI}{\dinit^{\scriptscriptstyle \mathrm{21cm}}}
\newcommand{\dinitJ}{\dinit^{\scriptscriptstyle \mathrm{joint}}}
\newcommand{\pxco}{P_{\scriptscriptstyle \mathrm{21cm}\times\mathrm{CO}}(k)}
\newcommand{\pxxi}{P_{\scriptscriptstyle \mathrm{21cm}}(k)}
\newcommand{\pCross}{P_{\scriptscriptstyle \mathrm{X}}(k)}
\newcommand{\pco}{P_{\scriptscriptstyle \mathrm{CO}}(k)}
\newcommand{\pinitJ}{P_{\scriptscriptstyle \mathrm{init}}^{\scriptscriptstyle \mathrm{joint}}(k)}
\newcommand{\pTrue}{P_{\scriptscriptstyle \mathrm{init}}^{\scriptscriptstyle \mathrm{true}}(k)}

\newcommand{\SBI}{SBI}
\newcommand{\MNRE}{MNRE}
\newcommand{\MCMC}{MCMC}
\newcommand{\NRE}{NRE}

\newcommand{\XXIonly}{\mbox{21-cm-only}}
\newcommand{\COonly}{\mbox{CO-only}}
\newcommand{\Joint}{\mbox{joint-inputs}}

\title[ML Reconstruction of Initial Density Field]{Tracing the Cosmic Origins: Machine Learning Reconstruction of the Primordial Density Field from EoR Observations}

\author[A. Saxena et al.]{
Anchal Saxena,$^{1}$\thanks{E-mail:
by.a.saxena@gmail.com}
P.\ Daniel Meerburg,$^{2}$
Guochao Sun,$^{3}$
Tzu-Ching Chang$^{4, 5}$
and Lluís Mas-Ribas$^{6}$
\\
$^{1}$Max-Planck-Insitut für Radioastronomie, Auf dem Hügel 69, D-53121 Bonn, Germany\\
$^{2}$Van Swinderen Institute, University of Groningen, Nijenborgh 4, 9747 AG Groningen, The Netherlands\\
$^{3}$CIERA and Department of Physics and Astronomy, Northwestern University, 1800 Sherman Ave, Evanston, IL 60201, USA\\
$^{4}$Jet Propulsion Laboratory, California Institute of Technology, 4800 Oak Grove Drive, Pasadena, CA 91109, USA\\
$^{5}$California Institute of Technology, 1200 E. California Blvd., Pasadena, CA 91125, USA\\
$^{6}$Department of Astronomy and Astrophysics, University of California, Santa Cruz, 1156 High Street, Santa Cruz, CA 95064, USA
}

\date{Accepted XXX. Received YYY; in original form ZZZ}

\pubyear{\the\year{}}

\begin{document}
\label{firstpage}
\pagerange{\pageref{firstpage}--\pageref{lastpage}}

\maketitle

\begin{abstract}
    Reconstructing the initial conditions of the Universe from late-time tracers would unlock cosmological information buried by non-linear structure formation and astrophysics. We reconstruct the initial density field corresponding to $z=300$, with its linear amplitude extrapolated to $z=0$, from simulated 21-cm and CO(1--0) line-intensity maps at $z=8$ generated with the semi-numerical code, \texttt{LIMFAST}. Using a three-dimensional U-Net, we reconstruct the initial conditions and evaluate their impact on cosmological parameter constraints. The two tracers probe complementary environments: 21-cm emission traces neutral, low-density regions of the intergalactic medium, while CO traces overdense, star-forming regions. We additionally train and evaluate the reconstruction on maps incorporating SKA1-Low-- and COMAP-ERA--like instrumental response and thermal noise. In the noiseless case, combining both tracers delivers the most accurate reconstruction across ionisation states, with the cross-correlation coefficient between the reconstructed and true  initial density fields, $|C(k)|\gtrsim0.90$ for $k\lesssim0.75\,{\rm Mpc}^{-1}$. With observational effects, small-scale information is degraded, but the joint reconstruction retains substantial large-scale coherence, with $|C(k)|\gtrsim0.70$ for $k\lesssim0.3\,{\rm Mpc}^{-1}$. To quantify the cosmological information gain, we perform simulation-based inference using power-spectrum summaries pre- and post-reconstruction. Across an ensemble of realisations, reconstruction yields the largest improvements for $\sigma_8$ and $n_{\rm s}$, with a median aggregate marginal compression of $1.17$ bits in the noiseless case, corresponding to a $56\%$ reduction in the factorised product of prior-normalised 68\% marginal widths. With observational effects, the corresponding median compression is $1.29$ bits. These results demonstrate that joint 21-cm--CO reconstruction can recover cosmological information beyond that captured by the pre-reconstruction summaries.
\end{abstract}



\section{Introduction}
The initial density perturbations generated during inflation seeded the present-day cosmic web, a hierarchy of structures from small scales such as galaxies and haloes to large-scale structures such as filaments and voids \citep{1992PhR...215..203M, 1996Natur.380..603B, Springel_2006}. As gravitational instability amplified these perturbations and structure formation progressed, the evolution became non-linear and coupled Fourier modes across different scales, redistributing cosmological information that had been explicit in the primordial two-point correlation function into higher-order correlations \citep{Rimes_2006, 2017PhRvD..96b3528C}. Consequently, constraints on cosmological parameters derived solely from late-time power spectra can be degraded \citep{Neyrinck_2006, 2021PhRvD.103b3506S, 2024CmPhy...7..130W, 2024PhRvD.109h3534H, frugte2025futureparameterconstraintsweak}. Reconstructing the initial density field can partially decouple these modes, recover otherwise lost information, and improve cosmological parameter constraints \citep{2013MNRAS.432..894J, 2013ApJ...772...63W, 2024JCAP...02..031F}.

Most reconstruction efforts to date have focused on low-redshift tracers ($z \sim 1$) available in galaxy surveys. Extending the same reconstruction from higher redshifts is particularly compelling. The earliest galaxies offer crucial insights into the initial phases of structure formation, and deep observations with the James Webb Space Telescope (JWST\footnote{\url{https://science.nasa.gov/mission/webb/}}) have already revealed populations at $z \lesssim 14$ \citep{2023MNRAS.526.2542C,  2023ApJS..265....5H, 2024ApJ...971...75M, 2024A&A...684A..75C, 2024MNRAS.532.1646K, 2024ApJ...969L...2F}. However, the narrow field of view of such surveys limits them to small regions of the sky, making them ill-suited to recover the large-scale initial density field. Line-intensity mapping (LIM) provides a complementary approach, delivering wide-field, three-dimensional measurements of the large-scale structure of the Universe \citep{2022A&ARv..30....5B}. In particular, the redshifted 21-cm line from neutral hydrogen enables tomographic mapping of the Cosmic Dawn and Epoch of Reionisation (CD-EoR) \citep{2015aska.confE...1K, mellema2015hi}.

The redshifted 21-cm signal is strongly shaped by astrophysical processes associated with the first stars and galaxies \citep{2001PhR...349..125B,2006PhR...433..181F,Pritchard_2012,Mu_oz_2022, 2025MNRAS.542.2292D}. X-ray heating and UV ionisation of the intergalactic medium (IGM) introduce additional mode coupling and imprint strong non-Gaussian features through spin-temperature and neutral-hydrogen fluctuations. Analyses with higher-order statistics such as the bispectrum \citep{2016MNRAS.458.3003S, 2020MNRAS.497.2941S, 2020MNRAS.499.5090M, 2021MNRAS.502.3800K, 2018MNRAS.476.4007M, 2020MNRAS.498.1480S} and trispectrum \citep{PhysRevD.77.103506, 2022JCAP...06..020F} have shown that these processes leave distinctive correlations in the 21-cm brightness-temperature field, complicating its interpretation. Beyond Fourier space statistics, multi-scale real-space approaches such as the wavelet scattering transform have also been shown to efficiently capture non-Gaussian information \citep{2022MNRAS.513.1719G, 2024ApJ...973...41Z, 2025PhRvD.112f3557S}. These considerations motivate reconstructing the initial density field, which mitigates non-linear mode coupling and restores an approximately linear, near-Gaussian field for which the power spectrum captures most of the cosmological information, thereby disentangling gravitational and astrophysical contributions.

The EoR 21-cm signal carries little to no information inside ionised regions, making complementary tracers essential. Emission from molecular and fine-structure lines such as CO(1--0) \citep{2011ApJ...741...70L, 2011ApJ...728L..46G, 2022ApJ...933..188B} and $[\CII]$ \citep{2012ApJ...745...49G, 2020ApJ...892...51C, 2024MNRAS.531.2958H} probes dense, star-forming galaxies that drive reionisation and therefore preferentially traces overdense regions of the matter density field. As a result, these tracers are statistically anti-correlated with the 21-cm signal on scales larger than the ionised bubbles and provide a complementary view of the ionisation morphology and its evolution \citep{Lidz_2011, 2021ApJ...909...51Z, 2024ApJ...975..222F}.

Building on this foundation, we reconstruct the initial matter density field from simulated 21-cm and CO line-intensity maps of the EoR using a three-dimensional U-Net. U-Nets \citep{2015arXiv150504597R} are well suited to this task because their encoder--decoder architecture with skip connections captures non-linear, multi-scale, and spatially complex relationships both within and across these tracers. U-Nets have already been applied in 21-cm cosmology for a wide variety of tasks, including foreground mitigation \citep{2021MNRAS.504.4716G,2024MNRAS.529.3684K, 2024arXiv240816814B, 2024PhRvD.109f3509S, Gao_2025}, segmentation of ionised and neutral regions \citep{Bianco_2021,2023arXiv230312065M}, and recovery of IGM parameters from 21-cm forest spectra \citep{patil2025efficientneutraligminferencenoisy}. Beyond 21-cm applications, U-Nets have proven effective for other non-linear problems in cosmology \citep{2021ApJ...907...44V}, including reconstruction of the initial density field from dark matter haloes \citep{Bottema_2025} and from the $z=0$ dark matter density field \citep{2024JCAP...02..031F}, with both approaches shown to tighten constraints on $\Lambda$CDM parameters and primordial non-Gaussianity.

Complementary to these machine-learning (ML) efforts, \citet{2024ApJ...965...31Z} presented an analytical reconstruction of the initial density field from the EoR 21-cm and CO line-intensity maps using a conjugate-gradient solver with fixed cosmological parameters. We extend this direction in two ways: (i) we allow cosmological parameters to vary within an ML-based framework and quantify the information gain by estimating constraints on cosmology pre- and post-reconstruction; and (ii) we incorporate a more realistic treatment of observational effects for both tracers. 

We consider the instrumental specifications of two upcoming facilities expected to deliver tomographic imaging of the EoR. For the 21-cm signal, we adopt SKA1-Low\footnote{\url{https://www.skao.int/en/explore/telescopes/ska-low}}, a low-frequency radio interferometer operating over 50--350\,MHz \citep{2019arXiv191212699B}, corresponding to $z \sim 27$--3. Its dense core and large collecting area provide the sensitivity required for arcminute-scale imaging of the 21-cm brightness-temperature field \citep{mellema2015hi,2015aska.confE...1K}. For CO(1--0), we base our forward model on the Carbon Monoxide Mapping Array Project (COMAP\footnote{\url{https://comap.caltech.edu/index.html}}). The current pathfinder phase operates a 10.4\,m antenna at the Owens Valley Radio Observatory (OVRO) in the 26--34\,GHz band \citep{2022ApJ...933..182C}, targeting CO(1--0) at $z \sim 2.4$--3.4 \citep{Chung_2022, 2024A&A...691A.336S}. Its planned EoR extension, COMAP-EoR, will deploy more 30 GHz receivers and add a 12--20\,GHz band, giving access to the CO(1--0) signal at $z \sim 4.8$--8.6 \citep{2022ApJ...933..182C, Breysse_2022}. However, the baseline design is not expected to reach the sensitivity required for tomographic CO(1--0) imaging during the EoR. We therefore adopt a more ambitious third-generation, hypothetical ``Extended Reionization Array'' (COMAP-ERA), as described in \citet{Breysse_2022}, which expands the number of dishes and total integration time, further increasing the sensitivity in both the 16 and 30 GHz bands for tomographic measurements.

For parameter inference, we adopt Simulation-Based Inference (\SBI\footnote{\url{https://simulation-based-inference.org}}), a modern likelihood-free paradigm \citep{Alsing_2019, Cranmer_2020, lueckmann2021benchmarking}. Classical likelihood-based methods (e.g.\ \MCMC, nested sampling) require an explicit, tractable likelihood and typically sample the full joint posterior over all parameters \citep{Handley_2015, Speagle_2020}. For high-dimensional, non-linear models, this is often intractable and computationally expensive. In contrast, \SBI\ learns the data likelihood implicitly from forward simulations and enables efficient inference of marginal posteriors. In this work, we use marginal neural ratio estimation (\MNRE) to learn likelihood-to-evidence ratios for the parameters of interest. This method provides fast, amortised estimates of the marginal posteriors and has been applied to a range of astrophysical and cosmological inference problems including CMB and large-scale structure analyses \citep{Cole_2022,Franco_Abell_n_2024}, 21-cm cosmology \citep{10.1093/mnras/stad2659,Saxena:2024rhu}, gravitational-wave inference \citep{bhardwaj2023peregrine, alvey2023simulationbased}, and Type Ia supernovae \citep{karchev2024sidereal}.

This article is organised as follows. Section \ref{sec:sims} describes the simulations and data products for the 21-cm and CO line-intensity maps. Section \ref{sec:reconstruction_framework} presents the U-Net--based reconstruction framework, including the network architecture, training, and optimisation. Section \ref{sec:inst_noise} models observational effects, including instrumental response and thermal noise. Section \ref{sec:sbi_setup} details our simulation-based inference setup. Section \ref{sec:results} reports reconstruction performance in noiseless and noisy settings and the resulting cosmological parameter constraints, including posterior-coverage checks and information-gain diagnostics. Section \ref{sec:summary} summarises our findings and outlines future prospects.  

\section{Simulations and Training Data}
\label{sec:sims}
\begin{figure*}
    \centering
    \includegraphics[width=\linewidth]{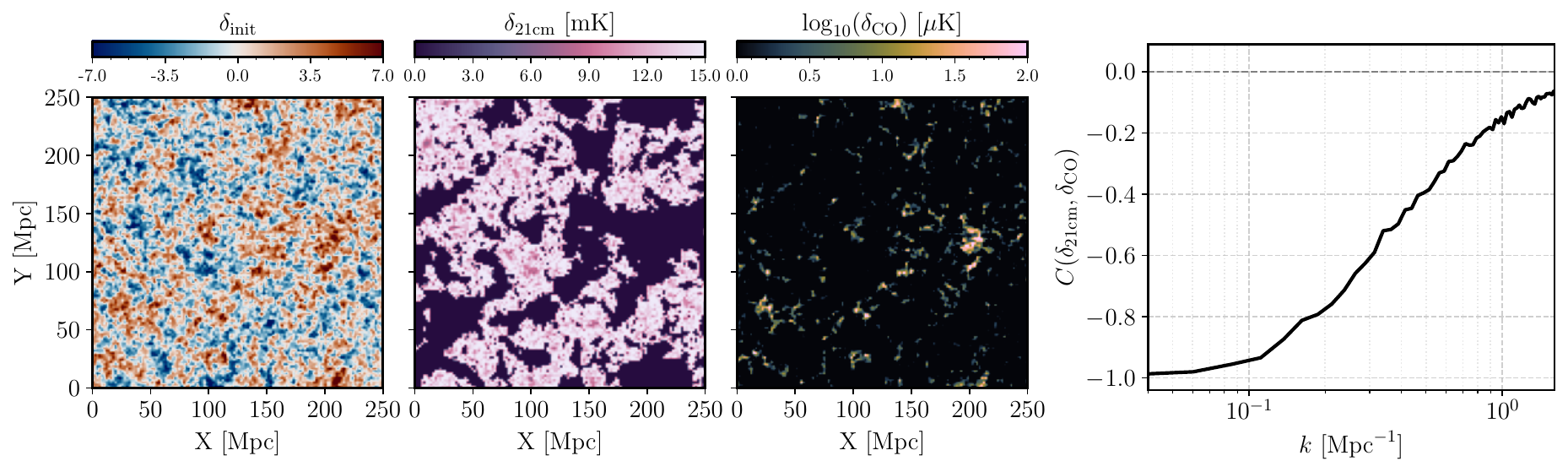}
    \caption{Representative 2D slices of the initial density field ($\dtrue$), whose amplitude is extrapolated to $z=0$, the 21-cm brightness-temperature field ($\dXXI$), and CO (1--0) line-intensity fluctuations ($\dCO$). The right panel shows the cross-correlation coefficient $C(\dXXI, \dCO)$, demonstrating the strong anti-correlation between the two tracers and their complementarity.}
    \label{fig:sample_data}
\end{figure*}
\begin{table*}
    \caption{Prior ranges for the cosmological parameters.}
    \label{tab:cosmo_params}
        \begin{tabular}{lcccccc}
        \hline
        Parameter & $\sigma_{8}$ & $h$ & $\Omega_{\rm m}$ & $\Omega_{\rm b}$ & $n_{\rm s}$\\
        \hline
        Prior & (0.70, 0.95) & (0.65, 0.71) & (0.25, 0.35) & (0.047, 0.049) & (0.90, 1.00) \\
        \hline
        \end{tabular}
\end{table*}
In this section, we present our simulations of the 21-cm and CO(1--0) line-intensity maps of the EoR. We used a semi-numerical code for simulating high-redshift galaxy formation and cosmic reionisation, \texttt{LIMFAST}\footnote{\url{https://github.com/lluism/LIMFAST}} \citep{2023ApJ...950...39M, 2023ApJ...950...40S, 2025ApJ...981...92S}. \texttt{LIMFAST} extends \texttt{21cmFAST} \citep{2007ApJ...669..663M, 2011MNRAS.411..955M} by incorporating self-consistent models of galaxy formation, evolution, chemical enrichment, and line emissivities for metal-line tracers within a unified cosmological framework.

In \texttt{21cmFAST}, large-scale density and velocity fields are generated from Gaussian initial conditions using Lagrangian perturbation theory \citep{1970A&A.....5...84Z, 1998MNRAS.299.1097S}. The collapsed fraction is then computed via the extended Press-Schechter formalism \citep{1993MNRAS.262..627L}, and the ionisation state of the IGM is computed with the excursion-set approach of \citet{2004ApJ...613....1F}, which compares the cumulative ionising-photon budget to the local neutral-hydrogen content. This yields three-dimensional realisations of the matter overdensity $\delta_{\rm b}(\boldsymbol{r},z)$, the neutral-hydrogen fraction $x_{\HI}(\boldsymbol{r},z)$, and the corresponding 21-cm brightness-temperature field $\delta T_{\rm b}(\boldsymbol{r},z)$, where $\boldsymbol{r}$ denotes the comoving spatial coordinate and $z$ the redshift,
\begin{equation}
    \begin{aligned}
    \delta T_{\rm b}(\boldsymbol{r},z)
       &= 27\,x_{\HI}(\boldsymbol{r},z)\,[1+\delta_{\rm b}(\boldsymbol{r},z)]\,f_z\,f_{\rm cosmo}\ \mathrm{mK},\\[3pt]
    {\rm where}\ f_z &\equiv \left[\frac{1+z}{10}\right]^{1/2},\
    f_{\rm cosmo} \equiv 
    \left[\frac{0.15}{\Omega_{\rm m} h^2}\right]^{1/2}
    \left[\frac{\Omega_{\rm b} h^2}{0.023}\right]\,,
    \end{aligned}
    \label{eq:tb}
\end{equation}
where we neglect redshift-space distortions and assume a saturated spin temperature, i.e.\ $T_{\rm S} \gg T_{\rm CMB}$, which is a reasonable approximation once the mean neutral fraction drops below $\langle x_{\HI}\rangle \lesssim 0.9$.

\texttt{LIMFAST} inherits the density, velocity, and ionisation fields from \texttt{21cmFAST} and forward-models the radiation from galaxies and the IGM. Galaxy evolution follows the quasi-equilibrium (``bathtub'') framework of \citet{2021MNRAS.500.3394F}, which self-consistently balances cosmological inflows, star-formation, and feedback-driven outflows to predict gas mass, stellar mass, and metallicity of galaxies as a function of halo mass and redshift. Stellar population synthesis models (e.g.\ \texttt{BPASS}) \citep{2017PASA...34...58E, 10.1093/mnras/sty1353} provide metallicity-dependent stellar spectral energy distributions (SEDs), while nebular line emissivities are computed with \texttt{CLOUDY}.

The luminosity of each halo is computed by integrating these emissivity grids, given the predicted gas density, metallicity, and local radiation field. \texttt{LIMFAST} integrates halo luminosities over the conditional halo mass function conditioned on the local overdensity, normalising sub-grid Press-Schechter counts to match Sheth-Tormen means. For neutral-ISM tracers such as $[\CII]$ and CO, the neutral-phase module ties emissivities to metallicity and interstellar radiation-field strength, allowing predictions that respond to star-formation law and feedback choices. For further details about \texttt{LIMFAST}, we refer the interested reader to \citet[][see also \citet{2025arXiv250907060S} for a simulation-based inference framework to infer galaxy-formation model parameters]{2023ApJ...950...39M}. 

\subsection{Data Products}
\label{sec:Data_products}
From this simulation suite, we construct a dataset of 600 coeval cubes at $z=8$ within a $(250\,\mathrm{Mpc})^3$ box discretised on a $128^3$ grid, corresponding to a voxel size $\simeq 1.95\,\mathrm{Mpc}$. To isolate sensitivity to cosmology, we vary cosmological parameters across these realisations while holding astrophysical parameters fixed. Specifically, we draw $\theta_{\rm cosmo} = \{\sigma_{\rm 8}, h, \Omega_{\rm m}, \Omega_{\rm b}, n_{\rm s}\}$ independently from flat priors within the intervals listed in Table~\ref{tab:cosmo_params} using a Latin hypercube sampling scheme, which ensures an efficient and approximately space-filling coverage of the cosmological parameter space. We further use independent random seeds for each realisation.

Each realisation has three co-registered fields: (i) the initial matter density ($\dtrue$) at $z=300$, (ii) the 21-cm brightness-temperature field ($\delta T_{\rm b}$) at $z=8$, hereafter denoted $\dXXI$, and (iii) the CO(1--0) line-intensity field ($\dCO$) at $z=8$. We note that although $\dtrue$ represents the initial conditions from which the subsequent density evolution is generated, \texttt{LIMFAST} stores this field using its linearly extrapolated $z=0$ normalisation. Throughout this work, we therefore use $\dinit$ to denote the linear initial density field extrapolated to the present day. Figure~\ref{fig:sample_data} shows representative slices of these fields together with the scale-dependent cross-correlation coefficient between the 21-cm and CO maps. The substantial anti-correlation illustrates that the two tracers probe complementary aspects of the EoR density and ionisation fields.

\begin{figure*}
    \centering
    \includegraphics[width=0.91\linewidth]{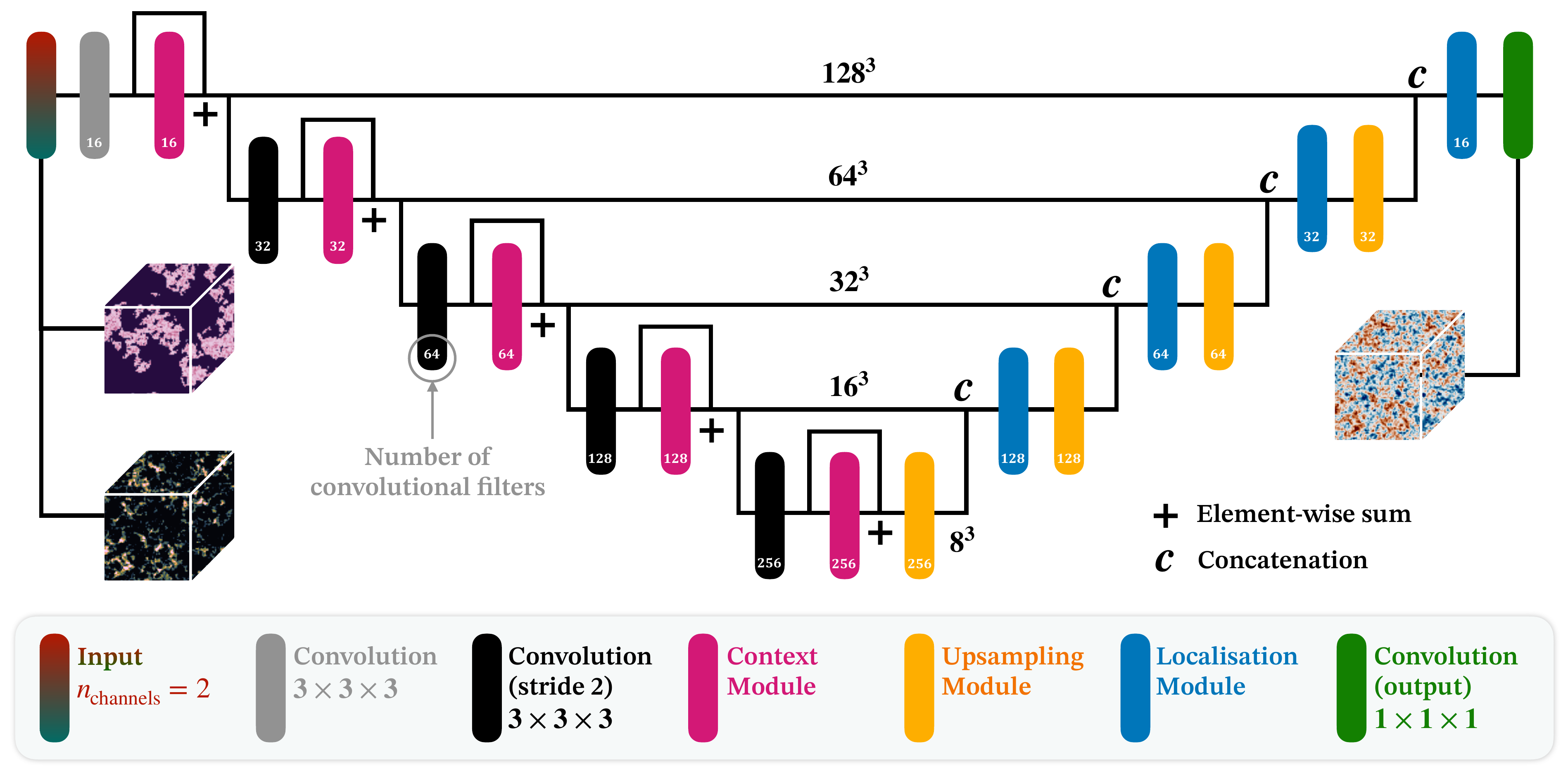}
    \caption{Schematic of the 3D residual U-Net used in this work. The encoder (left) compresses the input via stride-2 convolutions; the decoder (right) upsamples back to full resolution, and skip connections (horizontal links) pass high-resolution features forward.}
    \label{fig:uNet_arch}
\end{figure*}

The resulting dataset provides paired 3D volumes for supervised learning: the two-channel input $(\dXXI, \dCO)$ and the corresponding target ($\dtrue$). We partition the 600 realisations into 400, 100, and 100 samples for training, validation, and testing, respectively. Although 600 simulations may appear modest, each three-dimensional cube contains $\mathcal{O}(10^6)$ Fourier modes, providing ample statistical diversity. In practice, 3D U-Nets often train effectively on tens to a few hundred volumes when each volume is information-rich \citep{2024JCAP...02..031F, 2021MNRAS.504.4716G,Bottema_2025}. For computational efficiency, we focus on a single redshift ($z=8$); however, the framework naturally generalises to multi-redshift training, which should further improve performance by exploiting the redshift evolution of the two tracers.

\section{Reconstruction Framework}
\label{sec:reconstruction_framework}
We reconstruct the initial linear density field from simulated 21-cm and CO line-intensity cubes using a three-dimensional U-Net. Originally developed for biomedical image segmentation \citep{2015arXiv150504597R}, U-Nets are well suited to problems that involve the coupling of information across multiple spatial scales \citep{2021MNRAS.504.4716G, 2021ApJ...907...44V, 2024MNRAS.529.3684K, 2024arXiv240816814B, 2024PhRvD.109f3509S}.

Unlike standard convolutional neural networks (CNNs), whose receptive fields are local, a U-Net combines a down-sampling path that expands the receptive field and captures global context with an up-sampling path that restores full resolution while re-injecting small-scale features through skip connections. This multi-scale design has proved highly effective in astrophysical and cosmological applications, including reconstruction tasks closely related to ours \citep{2024JCAP...02..031F, Bottema_2025}.

\subsection{U-Net Architecture}
Our model follows the 3D residual U-Net architecture described by \citet{2021MNRAS.504.4716G}, adapted to our EoR data products. A schematic is shown in Figure~\ref{fig:uNet_arch}. The network operates on cubic fields defined on a $128\times128\times128$ grid, with a two-channel input (21-cm and CO) and a single-channel output (the reconstructed initial density field) at the same resolution. 

The encoder (down-sampling path) begins with a $3\times3\times3$ convolution followed by a \emph{context module} comprising two additional $3\times3\times3$ convolutional layers. To aid optimisation and prevent vanishing gradients, the output of the context module is added back to the output of the initial convolution through a \emph{residual connection} \citep{he2015deepresiduallearningimage}. All convolutions, except the final projection, are followed by instance normalisation and a leaky ReLU activation, which together stabilise training and improve the network's ability to represent complex, non-linear mappings \citep{Maas2013, 2015arXiv150500853X}. The encoder proceeds with a series of stride-2 convolutions that halve the spatial dimensions at each level. Each level mirrors the context module pattern described above. This process is repeated four times, resulting in feature maps of dimensions $8\times8\times8$ at the bottleneck. Intuitively, moving down this path increases the effective receptive field so that the model ``sees'' larger structures.

The decoder (up-sampling path) restores full spatial resolution via transposed convolutions that successively double the size of feature maps. At each level, features from the corresponding encoder stage are concatenated with the decoder features via \emph{skip connections}, preserving high-resolution details. The merged features are processed by a \emph{localisation module} consisting of two $3\times3\times3$ transposed convolutions. Finally, a $1\times1\times1$ convolution maps the output of the last decoder layer to a single-channel output without changing spatial resolution. 

The complete model consists of $\sim 2 \times 10^6$ trainable parameters and is implemented in \texttt{PyTorch}\footnote{\url{https://pytorch.org}} \citep{2019arXiv191201703P}. The implementation, named \texttt{CosmicRewind}, is publicly available on GitHub\footnote{\url{https://github.com/anchal-009/CosmicRewind}}.
\begin{figure*}
    \centering
    \includegraphics[width=0.98\linewidth]{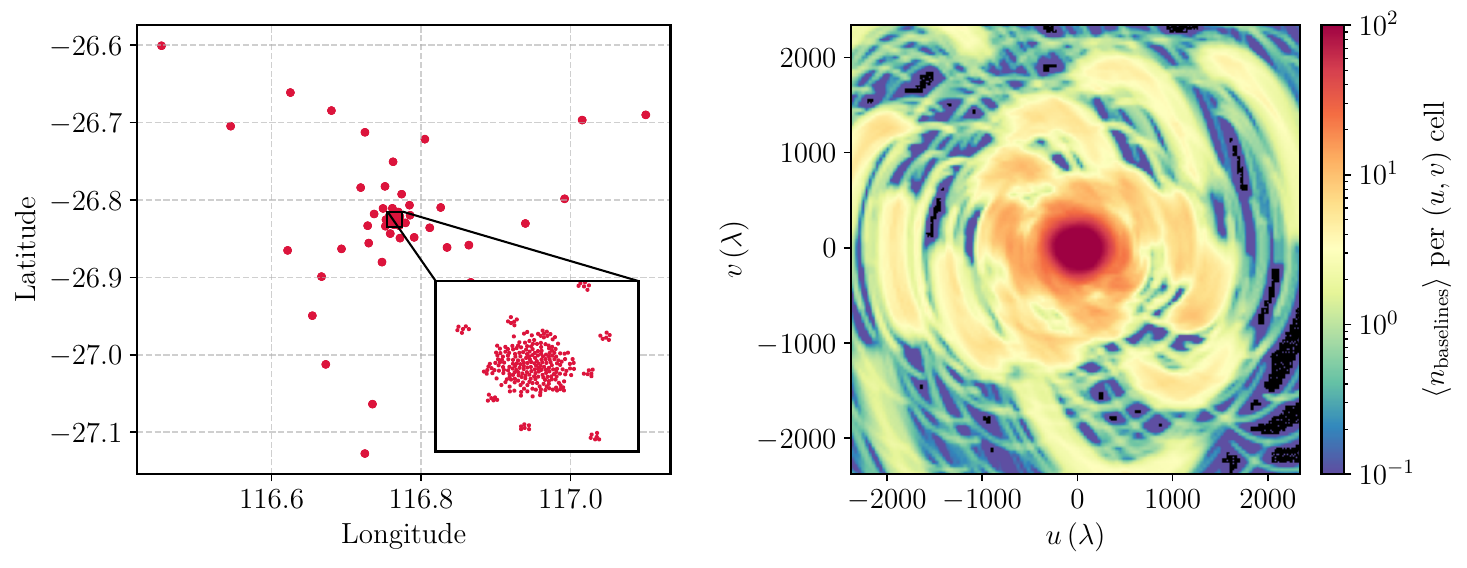}
    \caption{Left: Distribution of 512 SKA1-Low antenna stations used in our simulations. The inset shows the dense 1 km core containing 224 stations. Right: Corresponding six-hour uv-coverage for an observation centred at declination $-30^\circ$. The colour scale indicates the average number of baselines per $(u,v)$ cell, highlighting the dense sampling at short baselines that dominates the array sensitivity.}
    \label{fig:ska_uv_coverage}
\end{figure*}

\subsection{Training and Optimisation}
We train the network on the 400 realisations in the training set, using paired 21-cm and CO inputs at $z=8$ and the corresponding initial linear density field $\dtrue$, defined in Section~\ref{sec:Data_products}. All fields are generated on a $128^3$ grid.

We consider two training configurations. We first train a U-Net directly on the simulated noiseless 21-cm and CO fields. We then train a separate U-Net on the corresponding fields after applying the instrumental response and thermal-noise models described in Section~\ref{sec:inst_noise}. The reconstruction target is unchanged in both cases and corresponds to the same underlying $\dtrue$. Training directly on the observationally degraded fields allows the second network to learn the statistical mapping between noisy, resolution-limited tracer maps and the underlying initial conditions.

The objective of the network is to learn the mapping from the two tracer fields to the underlying initial density field by minimising the mean-squared error (MSE) between the reconstructed $\delta_{\rm init}^{\rm rec}(\boldsymbol{r})$ and the true initial density field $\delta_{\rm init}^{\rm true}(\boldsymbol{r})$:
\begin{equation}
    \ell_{\rm MSE}
    = \frac{1}{N_{\rm sims}} \sum_{i=1}^{N_{\rm sims}}
    \frac{1}{N_{\rm cells}} \sum_{\boldsymbol{r}}
    \left[ \delta_{\rm init,\, i}^{\rm true}(\boldsymbol{r}) - \delta_{\rm init,\,i}^{\rm rec}(\boldsymbol{r}) \right]^2,
    \label{eq:mse}
\end{equation}
where $N_{\rm sims}=400$ and $N_{\rm cells}=128^3$. The parameters of the network are optimised via stochastic gradient-based updates to minimise this loss. Through iterative weight updates, the model learns to capture the statistical relationship between the tracer fields and the underlying initial density fluctuations.

To mitigate over-fitting, we implement dropout layers \citep{10.5555/2627435.2670313} within the U-Net, which randomly deactivate a fraction of neurons during training to prevent the network from relying on any single pathway. We also employ early stopping based on the validation loss, halting training once no further improvement is observed on the 100 validation realisations. Performance is finally reported on the held-out 100 test realisations. Training on an NVIDIA V100 GPU typically converges within $\sim90\,{\rm min}$, after which a single forward pass produces a reconstruction in $\sim1\,{\rm s}$, making the approach computationally efficient for large-scale applications.

\subsection{Performance}
To evaluate the quality of reconstruction, we compute the cross-correlation coefficient between the reconstructed and true initial density fields. This metric quantifies the agreement between two fields across different scales in Fourier space and is defined as
\begin{equation}
    C(\delta_{\rm A}, \delta_{\rm B})(\boldsymbol{k}) = \frac{\langle\delta_{\rm A} (\boldsymbol{k})\delta_{\rm B}^{*}(\boldsymbol{k})\rangle}{\sqrt{P_{\rm A} (\boldsymbol{k})P_{\rm B}(\boldsymbol{k})}}\,,
\end{equation}
where $P_{\rm A}(\boldsymbol{k})$ and $P_{\rm B}(\boldsymbol{k})$ are the power spectra of fields $\delta_{\rm A}$ and $\delta_{\rm B}$, respectively, and $\langle \cdot \rangle$ denotes an average over modes within a $k$-bin. The absolute value, $\left|C(\delta_{\rm A}, \delta_{\rm B}; {\boldsymbol{k}})\right| \in [0,1]$, measures scale-dependent coherence, with unity indicating perfect coherence.

For the reconstructed field, we evaluate $|C(\dtrue, \dinitrec; k)|$. Because the cross-correlation coefficient measures coherence but is insensitive to an overall multiplicative difference in amplitude, we additionally compare the power spectra of the true and reconstructed initial density fields. We also report $|C(\dtrue,\dXXI;k)|$ and $|C(\dtrue,\dCO;k)|$ to benchmark the improvement in field-level coherence achieved by the reconstruction relative to the input tracer fields.

\section{Modelling Observational Effects}
\label{sec:inst_noise}
We simulate observational effects for both tracers by forward-modelling instrumental responses and thermal noise. For the 21-cm signal, we adopt SKA1-Low specifications\footnote{\href{https://www.skao.int/sites/default/files/documents/d17-SKA-TEL-SKO-0000557_01_-DesignConstraints-1.pdf}{SKA1-Low Design Constraints (PDF)}} \citep{2015aska.confE...1K, mellema2015hi}, while for the CO(1--0) line, we use specifications aligned with the COMAP-ERA project \citep{2022ApJ...933..182C, 2022ApJ...933..188B}.

\subsection{SKA1-Low}
For synthetic observations, we first simulate $uv$ tracks using the planned SKA1-Low antenna coordinates\footnote{\href{https://www.skao.int/sites/default/files/documents/d18-SKA-TEL-SKO-0000422_02_SKA1_LowConfigurationCoordinates-1.pdf}{SKA1-Low Antenna Coordinates (PDF)}}. The array consists of 512 stations, each with a diameter of $\sim$ 35--40\,m, with 224 non-overlapping stations randomly distributed within a compact $\sim 1$\,km core. The remaining stations are grouped into 48 clusters, each of which contains six randomly placed stations. These clusters are distributed in a three-armed spiral to $\sim 35$\,km radius. We simulate a six-hour track centred at declination $-30^\circ$ to obtain the $uv$ coverage. The station layout and the corresponding $uv$ coverage are shown in Figure~\ref{fig:ska_uv_coverage}.

Each pair of stations will record noise along with visibilities. We assume uncorrelated, zero-mean Gaussian noise \citep{10.1093/mnras/stw2494}, with rms per visibility 
\begin{equation}
    \sigma = \frac{\sqrt{2}k_{\rm B}T_{\rm sys}}{\epsilon A_{\rm D} \sqrt{\Delta \nu \Delta t}}\,,
\end{equation}
where $k_{\rm B}$ is the Boltzmann constant, $T_{\rm sys}$ is the system temperature, $\Delta \nu$ is the channel width, $\Delta t$ is the integration time per visibility, $A_{\rm D}$ is the physical area of each station, and $\epsilon$ accounts for the reduction in effective area above a characteristic frequency $\nu_c$:
\begin{equation}
    \epsilon = 
    \begin{cases}
    1, & \nu \leq \nu_{c},\\
    \left(\dfrac{\nu_{c}}{\nu}\right)^{2}, & \nu > \nu_{c}.
    \end{cases}
\end{equation}

We adopt the standard SKA1-Low system-temperature model
\begin{equation}
    T_{\rm sys}(\nu) = 60 \left(\frac{\nu}{300\,{\rm MHz}}\right)^{-2.55}\, {\rm K}\,.
\end{equation}
At the observed 21-cm signal frequency corresponding to $z = 8$ ($\nu \sim 158$\,MHz), this gives $T_{\rm sys} \sim 308$\,K. Our observing schedule assumes $t_{\scriptscriptstyle\rm obs}^{\scriptscriptstyle\rm day}$ = 6\,h tracks per day, for a total accumulated integration time of $t_{\scriptscriptstyle\rm obs}^{\scriptscriptstyle\rm tot}$ = 10,000\,h. In practice, we generate the $uv$ coverage for a single six-hour track, as shown in Figure~\ref{fig:ska_uv_coverage}, and then scale the thermal noise by $\sqrt{t_{\scriptscriptstyle\rm obs}^{\scriptscriptstyle\rm tot}/t_{\scriptscriptstyle\rm obs}^{\scriptscriptstyle\rm day}}$ to represent the total integration time. The adopted instrumental and observational parameters are summarised in Table~\ref{tab:inst_params}. 

We simulate the instrumental noise following the approaches of \citet{2018MNRAS.479.5596G} and \citet{10.1093/mnras/stw2494}, as implemented in \texttt{tools21cm}\footnote{\url{https://tools21cm.readthedocs.io}}. We first Fourier-transform the simulated 21-cm brightness-temperature cubes and sample them according to the simulated $uv$ coverage. Gaussian noise is then added to the sampled visibilities, scaled by the number of baselines contributing to each $uv$ cell. The noisy visibilities are then inverse-transformed to the image domain to obtain a thermal-noise–limited 21-cm line-intensity map.

\begin{table}
    \centering
    \caption{Instrumental and survey parameters adopted in this work.}
    \label{tab:inst_params}
    \begingroup
        \renewcommand{\arraystretch}{1.2}
        \begin{tabular}{l l |||||||||||||||||||||| l l}
            \hline
            \multicolumn{2}{c|}{\textbf{SKA1-Low (21-cm)}} &
            \multicolumn{2}{c}{\textbf{COMAP-ERA (CO)}} \\
            \hline
            \textit{Parameter} & \textit{Value} & \textit{Parameter} & \textit{Value} \\
            \hline
            $N_{\rm ant}$                 & 512                                 & $N_{\rm feeds}$ & 38 \\
            $T_{\rm sys}$                 & $308\,\mathrm{K}$ & $T_{\rm sys}$        & 20\,K \\
            $A_{\rm eff}$                 & $969\,\mathrm{m}^2$                  & $\Omega_{\rm field}$            & $4^\circ{}^2$\\
            $\Delta t$                    & 10\,s                                & $\delta\nu$         & 2\,MHz \\
            $t_{\rm obs}^{\rm day}$       & 6\,h                                 & $t_{\rm obs}$           & 57,000\,h \\
            $t_{\rm obs}^{\rm tot}$       & 10,000\,h                              & $\theta_{\rm FWHM}$    & 4\,{\rm arcmin}\\
            \hline
        \end{tabular}
    \endgroup
\end{table}

Most of the array sensitivity resides on short baselines. Accordingly, we restrict the effective baseline distribution to a maximum diameter of 2\,km around the SKA1-Low core. This naturally lowers the thermal-noise level at the expense of angular resolution. To match this resolution, we convolve the image cubes with a Gaussian kernel in the angular direction, with a full width at half maximum (FWHM) set by the 2\,km maximum baseline, and apply a top-hat filter of the same width along the line of sight. 

\begin{figure}
    \centering
    \includegraphics[width=\linewidth]{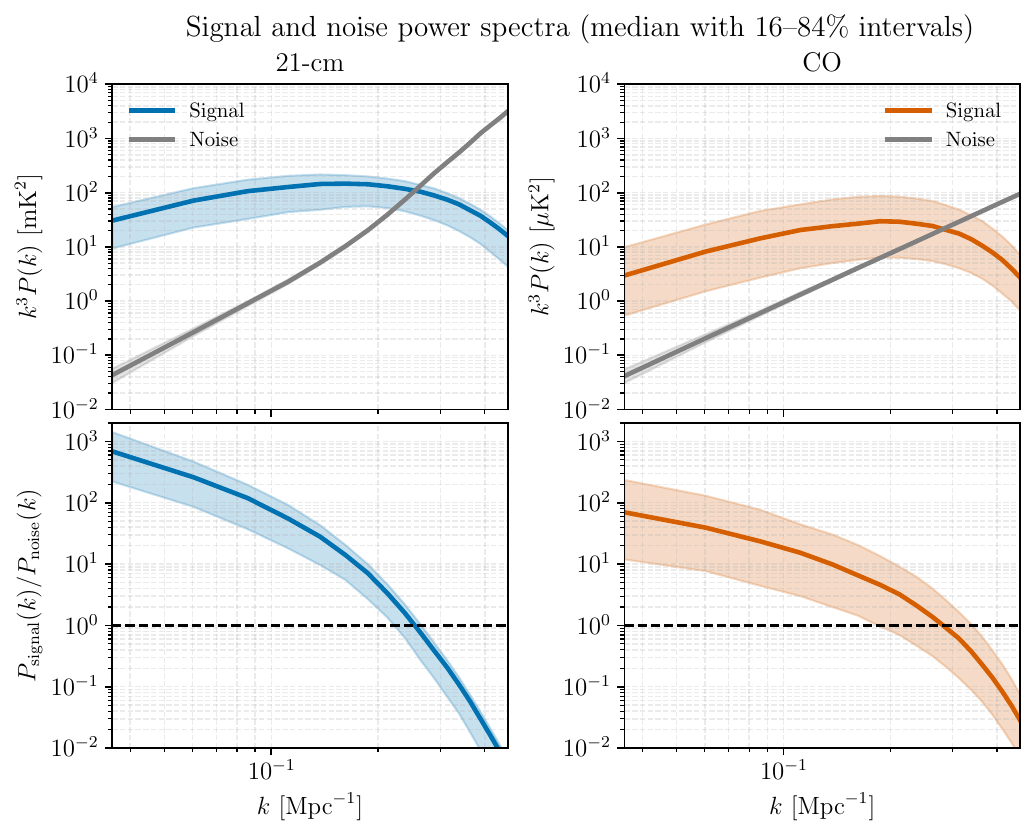}
    \caption{Signal and noise power spectra for the 21-cm (left) and CO(1--0) (right) intensity maps at $z=8$, computed from 500 simulated samples. The top panels show the median power spectra, with shaded regions indicating the 16--84\% range across the ensemble. The bottom panels show the ratio $P_{\rm signal}/P_{\rm noise}$, illustrating the scale dependence of the signal-to-noise.}
    \label{fig:snr_pk}
\end{figure}

\subsection{COMAP-ERA}
We forecast CO(1--0) intensity maps at $z = 8$ using the low-frequency (12--18\,GHz) sensitivity assumptions of the COMAP--ERA concept, i.e.\ a third-generation upgrade relative to COMAP--EoR \citep{2022ApJ...933..182C, Breysse_2022}. In the COMAP--EoR reference configuration, the current 30\,GHz pathfinder is augmented by two additional 30\,GHz receivers on existing 10.4\,m dishes and a new 16\,GHz receiver on an 18\,m dish, with five years of observing at $\sim\!1000$\,h\,yr$^{-1}$. This corresponds to $\sim\!7{,}000$ dish-hours at 16\,GHz. COMAP--ERA then assumes an expansion to ten dishes and a further five years of observing, increasing the total observation time to $\sim\!57{,}000$ dish-hours at 16\,GHz \citep{Breysse_2022}.

In the purely white-noise limit, the noise level in a map voxel is given as
\begin{equation}
    \sigma = \frac{T_{\rm sys}}{\sqrt{N_{\rm feeds}\,\delta\nu\,t_{\rm pix}}}\,,
\end{equation}
where $T_{\rm sys}$ denotes the system temperature, $N_{\rm feeds}$ is the number of independent feeds in a single dish, and $\delta\nu$ is the channel width. The factor $t_{\rm pix}$ represents the effective integration time per sky pixel, 
\begin{equation}
    t_{\rm pix} = t_{\rm obs}\left(\frac{\sigma_{\rm beam}^2}{\Omega_{\rm field}}\right)\,,
\end{equation}
where $t_{\rm obs}$ is the total observing time, and the ratio $\sigma_{\rm beam}^2/\Omega_{\rm field}$ accounts for the fraction of the total time a given sky pixel receives relative to the full survey area. The beam is modelled as a Gaussian with
\begin{equation}
    \sigma_{\rm beam} = \frac{\theta_{\rm FWHM}}{\sqrt{8\ln 2}}\,,
\end{equation}
where $\theta_{\rm FWHM}$ is the FWHM of the beam. To account for finite instrumental resolution, we smooth the maps in the angular direction with a Gaussian kernel of FWHM equal to $\theta_{\rm FWHM}$. Along the line of sight, we apply a top-hat filter with a width matched to the channel spacing. The adopted survey and instrument parameters are summarised in Table~\ref{tab:inst_params}.

To further illustrate the relative amplitude of the signal and noise across scales, we show in Figure~\ref{fig:snr_pk} the corresponding power spectra for the 21-cm and CO(1--0) intensity maps at $z =8$, computed across $500$ simulated samples. For each tracer, we show the median power spectrum, with the shaded regions indicating the $16$--$84\%$ range across the ensemble. The signal power spectra include the effect of finite angular resolution, while the noise spectra represent the instrumental noise level. For the CO signal, given the broad range of predicted $L_{\rm CO}$--$M_{\rm halo}$ relations in the literature, we adopt an optimistic scenario by rescaling the fiducial \texttt{LIMFAST} prediction by a factor of $100$, corresponding to the upper range of existing theoretical models (see, e.g., Figure~3 of \citealt{Breysse_2022}).

The figure illustrates how the signal and noise compare as a function of scale for the two tracers. In both cases, the signal power spectra decrease towards higher $k$ because finite angular resolution suppresses small-scale structure. The lower panels show the corresponding ratio $P_{\rm signal}/P_{\rm noise}$, which makes the scale dependence of the signal-to-noise more apparent and highlights the transition from signal-dominated to noise-dominated regimes. 

\section{Simulation-Based Inference Setup}
\label{sec:sbi_setup}
In this section, we give a brief overview of simulation-based inference (\SBI) \citep{Alsing_2019, Cranmer_2020, lueckmann2021benchmarking} and the specific implementation adopted in this work. Over the past five years, there has been rapid progress in the development and application of \SBI\ methods for data analysis \citep{Cranmer_2020}. At their core, these methods address a common question: \emph{can we perform robust Bayesian inference for a given generative model?}

Consider a forward generative model $p(\*x, \*\theta)$ that maps an underlying set of parameters $\*\theta$ to simulated data $\*x$. In a Bayesian sense, the forward model takes the form $p(\*x, \*\theta) = p(\*x\,|\,\*\theta)\,p(\*\theta)$, where $\*\theta$ is drawn from a prior distribution $p(\*\theta)$. This expression represents the key functionality of \SBI\ methods: running the forward model is equivalent to drawing samples from the simulated data likelihood $p(\*x\,|\,\*\theta)$. Scientifically, we are interested in inferring the posterior distribution $p(\*\theta\,|\,\*x)$ of the parameters $\*\theta$ given an observation $\*x$, which follows from Bayes' theorem,
\begin{equation}
    \label{eq:bayes}
    p(\*\theta\,|\,\*x) = \frac{p(\*x\,|\,\*\theta)}{p(\*x)}\, p(\*\theta)\,,
\end{equation}
where $p(\*x\,|\,\*\theta)$ is the likelihood of observing $\*x$ under parameters $\*\theta$, $p(\*\theta)$ is the prior distribution over the parameters, and $p(\*x)$ is the evidence or marginal likelihood of the data. 

Given the ability to sample from the forward model, \SBI\ methods typically construct the posterior using one of the three main approaches:
\begin{enumerate}
    \vspace{-0.5em}
    \item \textbf{Neural Posterior Estimation (NPE)} \citep{papamakarios2018fast, Tejero-Cantero2020, zeghal2022neural, dax2023group}, in which a flexible density estimator, such as a normalising flow, is trained to approximate the posterior $p(\*\theta\,|\,\*x)$ directly.

    \item \textbf{Neural Likelihood Estimation (NLE)} \citep{papamakarios2018fast, Alsing_2019, Lin_2023}, which focuses on learning an estimator for the likelihood $p(\*x\,|\,\*\theta)$, to be used subsequently with standard inference techniques such as {\small MCMC} or nested sampling.

    \item \textbf{Neural Ratio Estimation (NRE)} \citep{https://doi.org/10.5281/zenodo.5043706}, which aims to approximate the likelihood-to-evidence ratio $p(\*x\,|\,\*\theta) / p(\*x)$.
\end{enumerate}
In this work, we adopt NRE for parameter inference. In contrast to NPE and NLE, which learn conditional probability densities and therefore impose normalisation constraints that typically require specialised density-estimation architectures such as  normalising flows or mixture density networks, NRE learns the likelihood-to-evidence ratio and does not require explicit probability normalisation. This relaxes architectural constraints on the neural network and allows the use of simpler network architectures which we employ throughout this work.

\subsection{Neural Ratio Estimation}
In \SBI, information about the likelihood is \emph{implicitly} accessed through a stochastic simulator that maps \(\*\theta\mapsto\*x\). In neural ratio estimation (NRE), we build on this setup by generating a set of simulated pairs $\{(\*x^1,\*\theta^1), (\*x^2,\*\theta^2), \ldots\}$, where each $\*\theta^i$ is typically drawn from the prior. These pairs therefore follow the joint distribution $p(\*x,\*\theta)$ and are used to train a neural network to approximate the likelihood-to-evidence ratio for the parameters of interest \citep{hermans2020likelihoodfree, pmlr-v119-hermans20a, https://doi.org/10.48550/arxiv.2002.03712}. From equation~(\ref{eq:bayes}), this ratio, denoted $r(\*x, \*\theta)$, can be written as
\begin{equation}
    r(\*x, \*\theta) \equiv \frac{p(\*x\,|\,\*\theta)}{p(\*x)} = \frac{p(\*\theta\,|\,\*x)}{p(\*\theta)} = \frac{p(\*x, \*\theta)}{p(\*x)p(\*\theta)}\,.
\end{equation}
Thus, $r(\*x, \*\theta)$ is the ratio of the joint probability density $p(\*x, \*\theta)$ to the product of marginal densities $p(\*x)\,p(\*\theta)$. 

To estimate $r(\*x, \*\theta)$, we train a binary classifier $d_{\*\phi}(\*x, \*\theta)$ to distinguish between pairs drawn from the joint distribution and pairs constructed by independently drawing from the marginals. We assign a label $y=1$ to jointly drawn pairs $\*x, \*\theta \sim p(\*x, \*\theta)$ and $y=0$ to marginally drawn pairs $\*x, \*\theta \sim p(\*x)p(\*\theta)$. The trained classifier approximates the probability that a sample-parameter pair ($\*x, \*\theta$) was drawn jointly ($y=1$), i.e., \
\begin{equation*}
    \begin{split}
        d_{\*\phi} (\*x, \*\theta) &\approx p(y=1\,|\,\*x, \*\theta) \\
        &= \frac{p(\*x, \*\theta\,|\,y=1) p(y=1)}{p(\*x, \*\theta\,|\,y=1) p(y=1) + p(\*x, \*\theta\,|\,y=0) p(y=0)} \\
        &= \frac{p(\*x, \*\theta)}{p(\*x, \*\theta) + p(\*x) p(\*\theta)}\,,
    \end{split}
\end{equation*}
where we adopt balanced class priors, $p(y=0) = p(y=1) = 1/2$. We optimise the classifier parameters $\*\phi$ by minimising the binary cross-entropy loss
\begin{equation}
    - \iint \left[p(\*x, \*\theta) \ln d_\phi (\*x, \*\theta) + p(\*x)p(\*\theta) \ln \{1 - d_\phi (\*x, \*\theta)\}\right] \dd{\*x} \dd{\*\theta}\,,
\end{equation}
using stochastic gradient descent. In practice, $d_{\*\phi}(\*x, \*\theta)$ is implemented as a fully connected neural network with a few hidden layers. 

After training, the classifier output satisfies
\begin{equation}
    d_{\*\phi} (\*x, \*\theta) \approx \frac{p(\*x, \*\theta)}{p(\*x, \*\theta) + p(\*x) p(\*\theta)} = \frac{r(\*x, \*\theta)}{r(\*x, \*\theta) + 1}\,,
\end{equation}
which we then invert to obtain
\begin{equation}
    \label{eq:r_inversion}
    r(\*x,\*\theta)
    \approx \frac{d_{\*\phi}(\*x,\*\theta)}
    {1-d_{\*\phi}(\*x,\*\theta)}\,.
\end{equation}
Substituting into equation~(\ref{eq:bayes}) yields an estimator for the posterior,
\begin{equation}
    \label{eq:posterior_est}
    p(\*\theta\,|\,\*x)
    \approx r(\*x,\*\theta)\,p(\*\theta)\,.
\end{equation}

We are often interested in marginal posteriors rather than the full joint posterior. Instead of learning the full joint and marginalising afterwards, \NRE\ can be adapted to target these marginals directly by omitting nuisance parameters from the network inputs, an extension referred to as marginal neural ratio estimation (\MNRE). In this work, we use \MNRE\ as implemented in the software package \texttt{swyft}\footnote{\url{https://swyft.readthedocs.io/en/v0.3.2/}} \citep{Miller2022}.
\begin{figure}
    \centering
    \includegraphics[width=\linewidth]{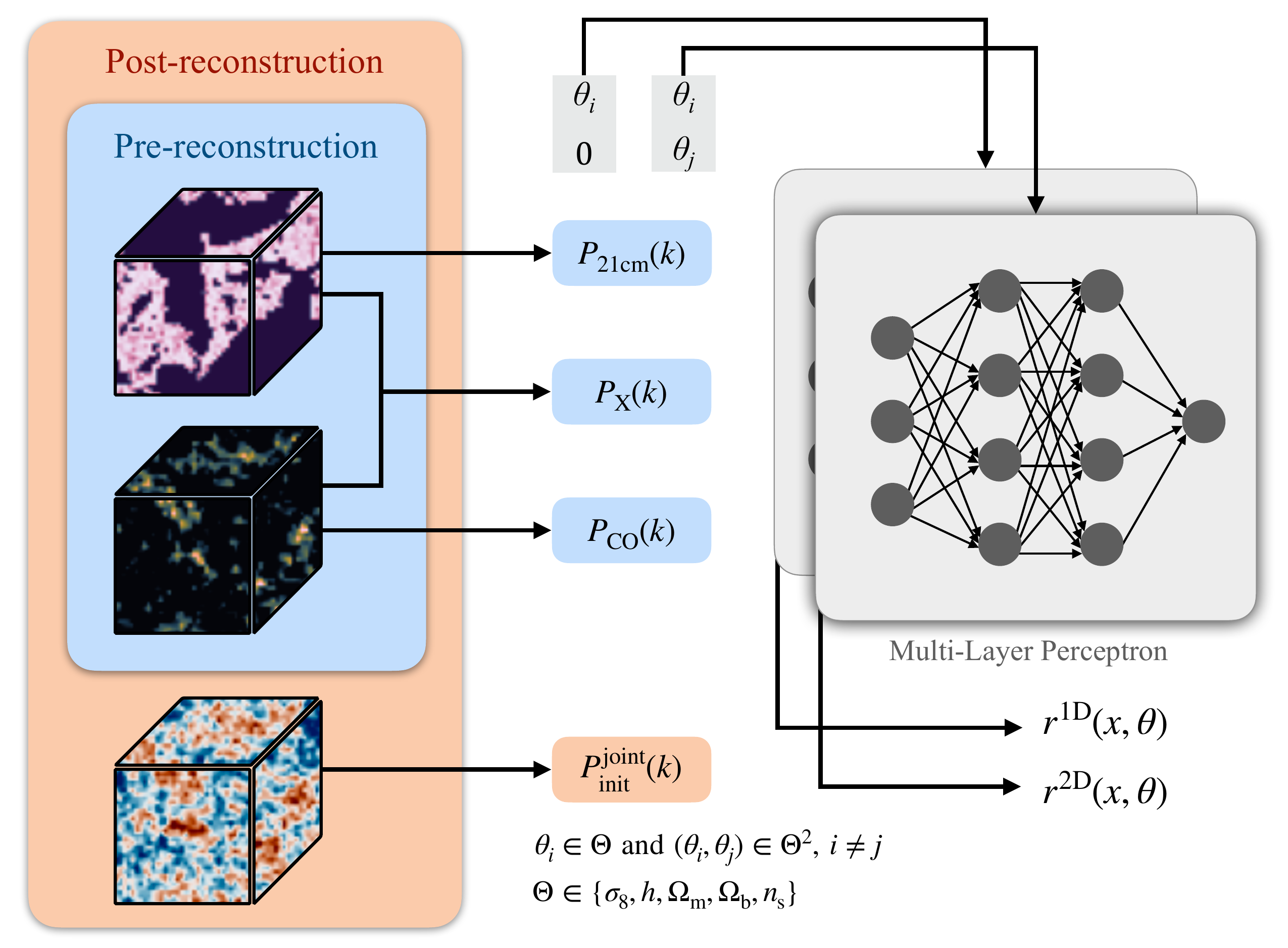}
    \caption{Schematic overview of the \SBI\ workflow. Pre-reconstruction summaries comprise the 21-cm and CO auto-power spectra, $\pxxi$ and $\pco$, together with their cross-power spectrum, $\pCross$, while post-reconstruction summaries additionally include the power spectrum of the jointly reconstructed initial density field, $\pinitJ$. These summaries are fed to a multi-layer perceptron that estimates likelihood-to-evidence ratios for the 1D and 2D marginal posteriors of the parameters $\theta_i \in \{\sigma_8,h,\Omega_{\rm m},\Omega_{\rm b},n_{\rm s}\}$.}
    \label{fig:sbi_network}
\end{figure}

\subsubsection{Power-spectrum summary sets}
\label{sec:summary_sets}
To apply this framework, we construct several power-spectrum summary sets that allow us to distinguish the information contained in the two-point cross-correlation of the tracers from the additional information supplied by the reconstructed initial density field. We first define an auto-power-spectrum-only summary,
\begin{equation}
    \mathcal{S}_{\rm auto}
    =
    \left\{
    \pxxi,\,
    \pco
    \right\}.
    \label{eq:summary_auto}
\end{equation}
We then include the 21-cm--CO cross-power spectrum, $\pCross \equiv \pxco$, to define our fiducial pre-reconstruction summary,
\begin{equation}
    \mathcal{S}_{\rm pre}
    =
    \left\{
    \pxxi,\,
    \pco,\,
    \pCross
    \right\}.
    \label{eq:summary_pre}
\end{equation}
The corresponding post-reconstruction summary is
\begin{equation}
    \mathcal{S}_{\rm post}
    =
    \left\{
    \pxxi,\,
    \pco,\,
    \pCross,\,
    \pinitJ
    \right\},
    \label{eq:summary_post}
\end{equation}
where $\pinitJ$ is the power spectrum of the initial density field reconstructed jointly from the 21-cm and CO maps. Figure~\ref{fig:sbi_network} provides a schematic overview of the pre- and post-reconstruction summary sets and their use in the SBI framework. Including $\pCross$ in $\mathcal{S}_{\rm pre}$ ensures that the two-point cross-correlation between the co-registered tracers is already included before $\pinitJ$ is introduced. The comparison $\mathcal{S}_{\rm pre}\rightarrow\mathcal{S}_{\rm post}$ therefore quantifies the incremental cosmological information provided by the power spectrum of the reconstructed initial density field beyond the selected auto- and cross-power spectra of the two tracers.

For reference, we additionally consider an idealised benchmark,
\begin{equation}
    \mathcal{S}_{\rm true}
    =
    \left\{
    \pxxi,\,
    \pco,\,
    \pCross,\,
    \pTrue
    \right\},
    \label{eq:summary_true}
\end{equation}
in which $\pinitJ$ is replaced by the power spectrum of the true initial density field, $\pTrue$.

\begin{figure*}
    \centering
    \includegraphics[width=0.875\linewidth]
    {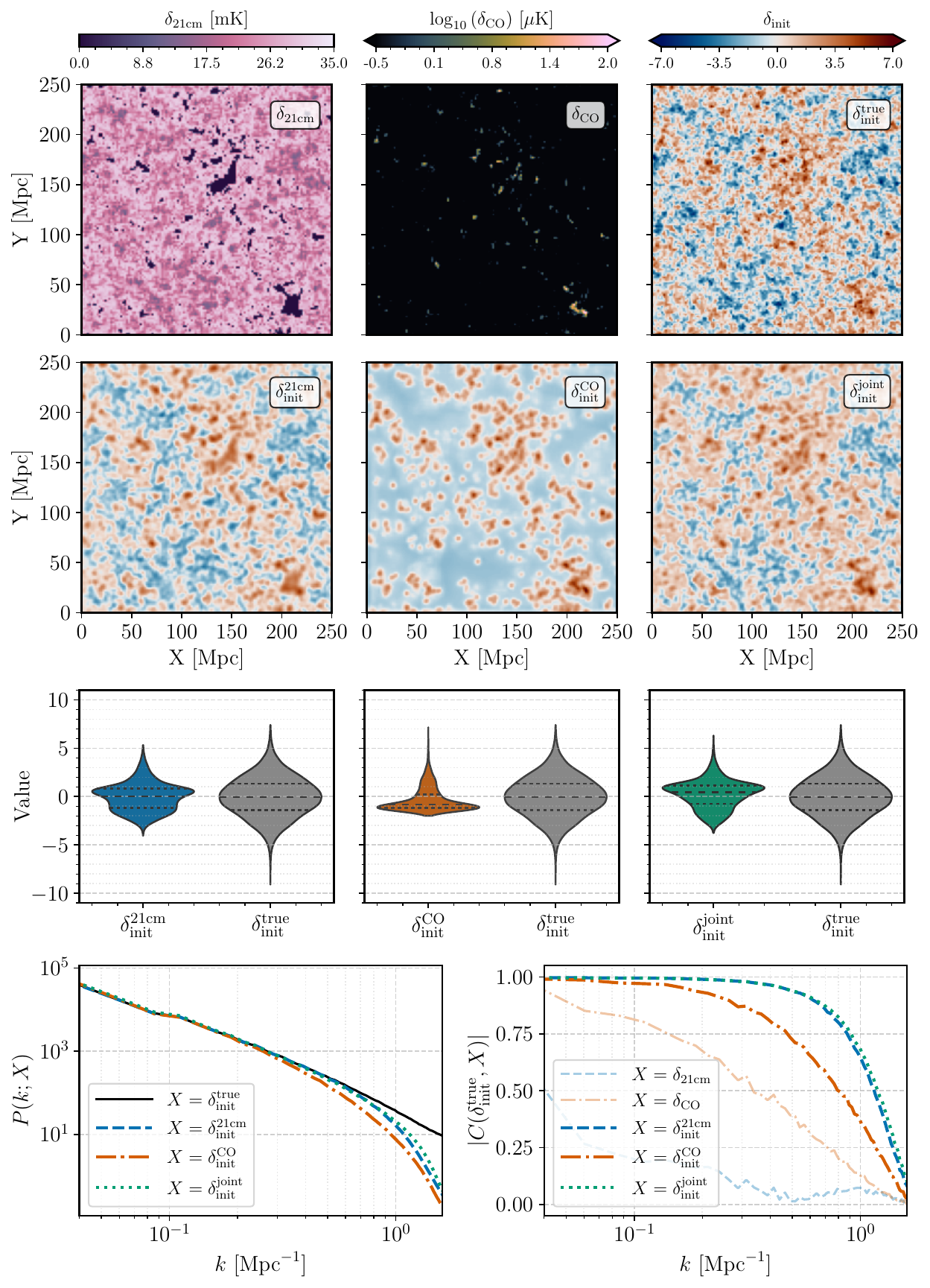}
    \caption{Reconstruction of the initial density field, expressed using its linear amplitude extrapolated to $z=0$, for a highly neutral sample with $(\sigma_8, h, \Omega_{\rm m}, n_{\rm s})$ = (0.717, 0.7, 0.252, 0.902) with $\langle x_{\HI}\rangle \approx 0.83$. The top row shows the input fields: ($\dXXI$ and $\dCO$) at $z \sim 8$ and ground truth: $\dtrue$. The second row shows the reconstructed initial density field when using 21-cm only, CO only, and joint maps as inputs to the U-Net. The third row shows the one-point probability density functions (PDFs) of the reconstructed and true initial density fields. The bottom row shows scale-dependent statistics of the reconstruction: the power spectra of the true and reconstructed initial density fields (left), and the cross-correlation between the true and reconstructed fields, $|C(\dtrue, X)|$ (right).}
   \label{fig:summary_plot_neutral}
\end{figure*}

\subsection{Information Gain from Reconstruction}
\label{sec:information_gain}
To quantify the cosmological information gained by adding different summary statistics, we compare the marginal posterior constraints obtained from the summary sets defined in Section~\ref{sec:summary_sets}. For each cosmological parameter $\theta_i$, we characterise the marginal posterior using the width $w_i$ of the 68\% highest-density credible interval and the Kullback--Leibler (KL) divergence between the posterior and the prior,
\begin{equation}
    {\rm KL}_i
    =
    \int
    p(\theta_i \,|\, {\*x})
    \log_2
    \left[
    \frac{p(\theta_i \,|\, {\*x})}
         {p(\theta_i)}
    \right]
    {\rm d}\theta_i ,
    \label{eq:kl_divergence}
\end{equation}
where $p(\theta_i \,|\, {\*x})$ is the marginal posterior inferred with MNRE and $p(\theta_i)$ is the corresponding prior. The use of a base-2 logarithm expresses ${\rm KL}_i$ in bits, with larger values indicating a posterior that is more informative relative to the prior.

For two summary sets $\mathcal{S}_A$ and $\mathcal{S}_B$, we define the change in marginal information as
\begin{equation}
    \Delta {\rm KL}_i(A\rightarrow B)
    =
    {\rm KL}_i(\mathcal{S}_B)
    -
    {\rm KL}_i(\mathcal{S}_A).
    \label{eq:delta_kl_general}
\end{equation}
Our primary reconstruction comparison is $\mathcal{S}_{\rm pre}\rightarrow\mathcal{S}_{\rm post}$, for which
\begin{equation}
    \Delta {\rm KL}_i^{\rm rec}
    =
    {\rm KL}_i(\mathcal{S}_{\rm post})
    -
    {\rm KL}_i(\mathcal{S}_{\rm pre}).
    \label{eq:delta_kl_rec}
\end{equation}
Because $\mathcal{S}_{\rm pre}$ and $\mathcal{S}_{\rm post}$ both contain $\pxxi$, $\pco$, and $\pCross$, this comparison measures the incremental marginal information obtained after adding $\pinitJ$ to the auto- and cross-power spectra of the two tracers. We also evaluate $\mathcal{S}_{\rm auto}\rightarrow\mathcal{S}_{\rm pre}$ to quantify the information added by the 21-cm--CO cross-power spectrum itself.

We quantify changes in the 68\% credible width using
\begin{equation}
I_{w,i}(A\rightarrow B)
=
100\left(
1-\frac{w_i^B}{w_i^A}
\right)\% ,
\label{eq:width_improvement}
\end{equation}
where $w_i^A$ and $w_i^B$ are the marginal credible widths obtained from $\mathcal{S}_A$ and $\mathcal{S}_B$, respectively. Positive values of $I_{w,i}$ therefore correspond to narrower marginal constraints after adding the new summary statistic, whereas negative values indicate broader credible intervals.

In addition to these parameter-by-parameter diagnostics, we define a factorised measure of the overall marginal posterior compression. For a summary configuration $c$, we first normalise each 68\% credible width by the corresponding prior width,
\begin{equation}
    r_i^{(c)}
    =
    \min\left(
    \frac{w_{i,c}}{\Delta\theta_i^{\rm prior}},\,1
    \right),
    \label{eq:normalized_width}
\end{equation}
where $\Delta\theta_i^{\rm prior}$ is the width of the prior for parameter $\theta_i$. The minimum in equation~(\ref{eq:normalized_width}) caps prior-limited constraints at unity. We then define the factorised marginal-volume proxy
\begin{equation}
    V_{\rm marg}^{(c)}
    =
    \prod_i r_i^{(c)},
    \label{eq:marginal_volume}
\end{equation}
and the corresponding marginal-compression statistic
\begin{equation}
    C_{\rm marg}^{(c)}
    =
    -\log_2 V_{\rm marg}^{(c)}.
    \label{eq:marginal_compression}
\end{equation}
The change in marginal compression between two summary
configurations is therefore
\begin{equation}
    \Delta C_{\rm marg}(A\rightarrow B)
    =
    C_{\rm marg}^{(B)}
    -
    C_{\rm marg}^{(A)}
    =
    \log_2
    \left(
    \frac{V_{\rm marg}^{(A)}}
         {V_{\rm marg}^{(B)}}
    \right).
    \label{eq:delta_marginal_compression}
\end{equation}
Positive values of $\Delta C_{\rm marg}$ indicate a reduction in the product of the prior-normalised marginal credible widths. In this definition, $\Delta C_{\rm marg}=1$ bit corresponds to a factor of two reduction in this factorised marginal-volume proxy. We emphasise that $C_{\rm marg}$ is constructed from the product of one-dimensional marginal credible widths and is therefore not a measure of the full joint posterior volume; in particular, it does not account for correlations between cosmological parameters.

Although the MNRE is trained to infer all five cosmological parameters $\{\sigma_8,h,\Omega_{\rm m},\Omega_{\rm b},n_{\rm s}\}$, we focus the quantitative information-gain analysis on $\{\sigma_8,h,\Omega_{\rm m},n_{\rm s}\}$. The marginal posterior of $\Omega_{\rm b}$ remains substantially prior limited for the summary statistics considered here. Consequently, changes in its credible width do not provide a reliable measure of reconstruction-induced information gain.

\subsection{Posterior Calibration Diagnostics}
In principle, one could validate the simulation-based inference by comparison with traditional sampling-based methods (e.g.\ MCMC, nested sampling), but such approaches are computationally expensive in our setting. Instead, we employ a complementary diagnostic to cross-validate and assess the statistical calibration of the inferred posteriors by means of an expected coverage test \citep{hermans2020likelihoodfree, lemos2023samplingbased}. This test measures whether Bayesian credible regions of nominal credibility $1-\alpha$ (where $\alpha$ denotes the nominal posterior error rate) achieve their expected coverage under repeated draws from the joint distribution $p(\*x, \*\theta)$.

Concretely, we consider a set of $N$ independent realisations $(\*x_j, \*\theta_j^*) \sim p(\*x, \*\theta)$, where $j=1,\ldots,N$. For each observation $\*x_j$, we compute the estimated posterior $\hat p(\*\theta\,|\,\*x_j)$ and define the corresponding $1-\alpha$ highest posterior density region (HPDR), denoted $\Omega_{\hat p(\*\theta|\*x_j)} (1-\alpha)$, which contains a fraction $1-\alpha$ of the posterior probability mass. We then record whether the ground-truth $\*\theta_j^*$ lies within this region. Averaging this indicator over many realisations gives an estimate of the empirical coverage probability for the nominal credibility level $1-\alpha$,
\begin{equation}
    \begin{aligned}
        1 - \hat{\alpha} &\equiv \mathds{E}_{(\*x, \*\theta) \sim p(\*x, \*\theta)}\mathds{1}\left[\*\theta \in \Omega_{\hat p (\*\theta|\*x)}(1-\alpha)\right] \\
        &\approx \frac{1}{N} \sum_{j=1}^N \mathds{1} \left[ \*\theta_j^* \in \Omega_{\hat p (\*\theta|\*x)} (1-\alpha) \right]\,,
    \end{aligned}
\end{equation}
where $\mathds{1}$ denotes the indicator function. In the limit where the estimated posterior $\hat p (\*\theta\,|\,\*x)$ approaches the true posterior $p(\*\theta\,|\,\*x)$, $1-\hat\alpha \approx 1-\alpha$. To emphasise deviations at high credibility, we reparameterise the nominal credibility level in terms of the Gaussian-equivalent quantile $q$, defined as
\begin{equation}
q=\Phi^{-1}(1-\alpha/2),
\end{equation}
where $\Phi^{-1}$ is the inverse cumulative distribution function of the standard normal distribution. We similarly define the empirical quantile as $\hat q=\Phi^{-1}(1-\hat\alpha/2)$. In this representation, the familiar Gaussian-equivalent $1\sigma$, $2\sigma$, and $3\sigma$ credibility levels correspond to $q=1$, $2$, and $3$, respectively. In a $\hat q$--$q$ coverage plot, agreement with the diagonal indicates proper calibration, while deviations indicate over- or under-coverage.

\section{Results}
\label{sec:results}
\begin{figure*}
    \centering
    \includegraphics[width=0.875\linewidth]{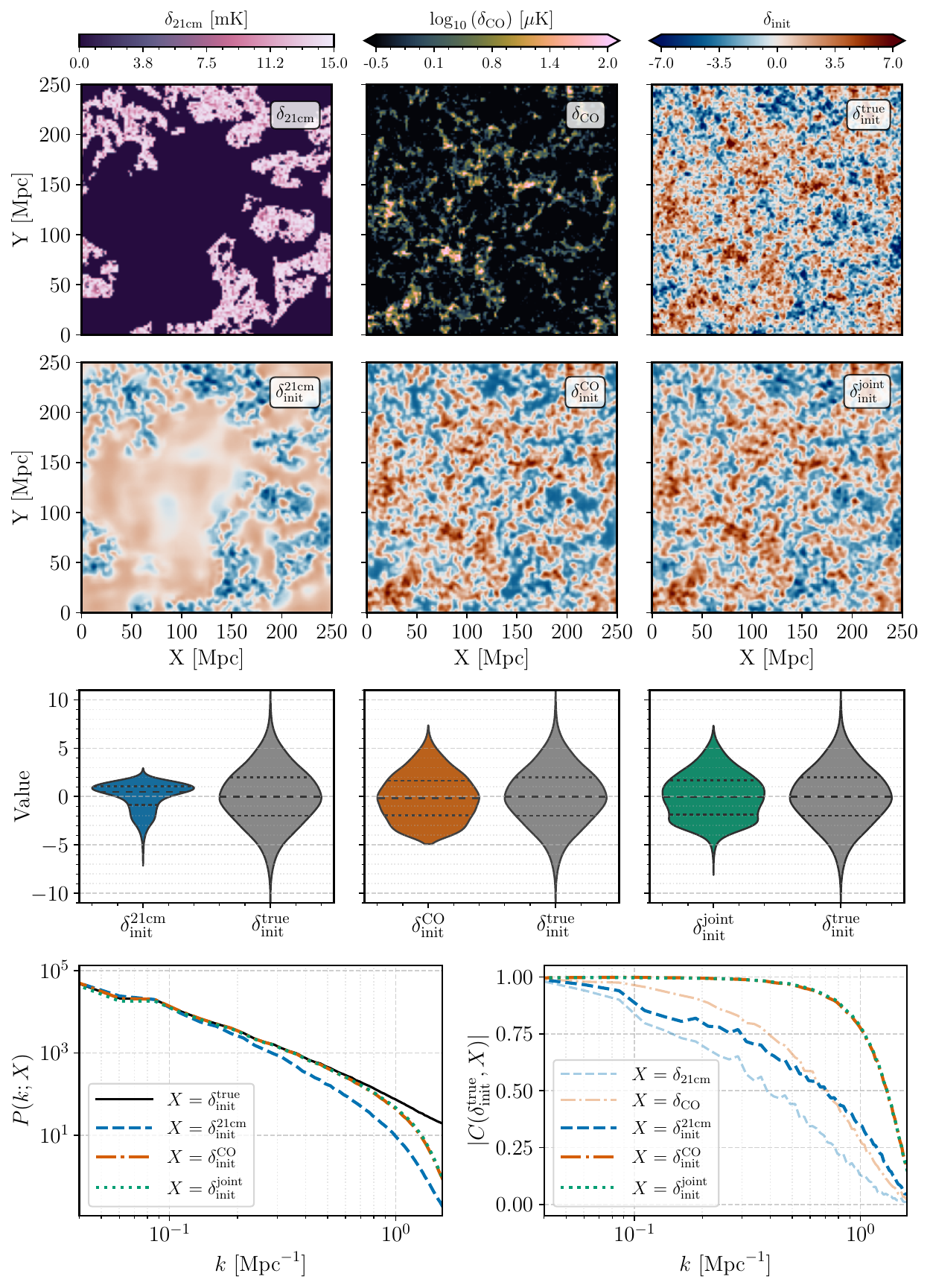}
    \caption{Reconstruction of the initial density field, expressed using its linear amplitude extrapolated to $z=0$, for a highly ionised sample with $(\sigma_8, h, \Omega_{\rm m}, n_{\rm s})$ = (0.917, 0.697, 0.291, 0.944) with $\langle x_{\HI}\rangle \approx 0.12$. The top row shows the input fields: ($\dXXI$ and $\dCO$) at $z \sim 8$ and ground truth: $\dtrue$. The second row shows the reconstructed initial density field when using 21-cm only, CO only, and joint maps as inputs to the U-Net. The third row shows the one-point probability density functions (PDFs) of the reconstructed and true initial density fields. The bottom row shows scale-dependent statistics of the reconstruction: the power spectra of the true and reconstructed initial density fields (left), and the cross-correlation between the true and reconstructed fields, $|C(\dtrue, X)|$ (right).}
    \label{fig:summary_plot_ionized}
\end{figure*}
\begin{figure*}
    \centering
    \includegraphics[width=\linewidth]{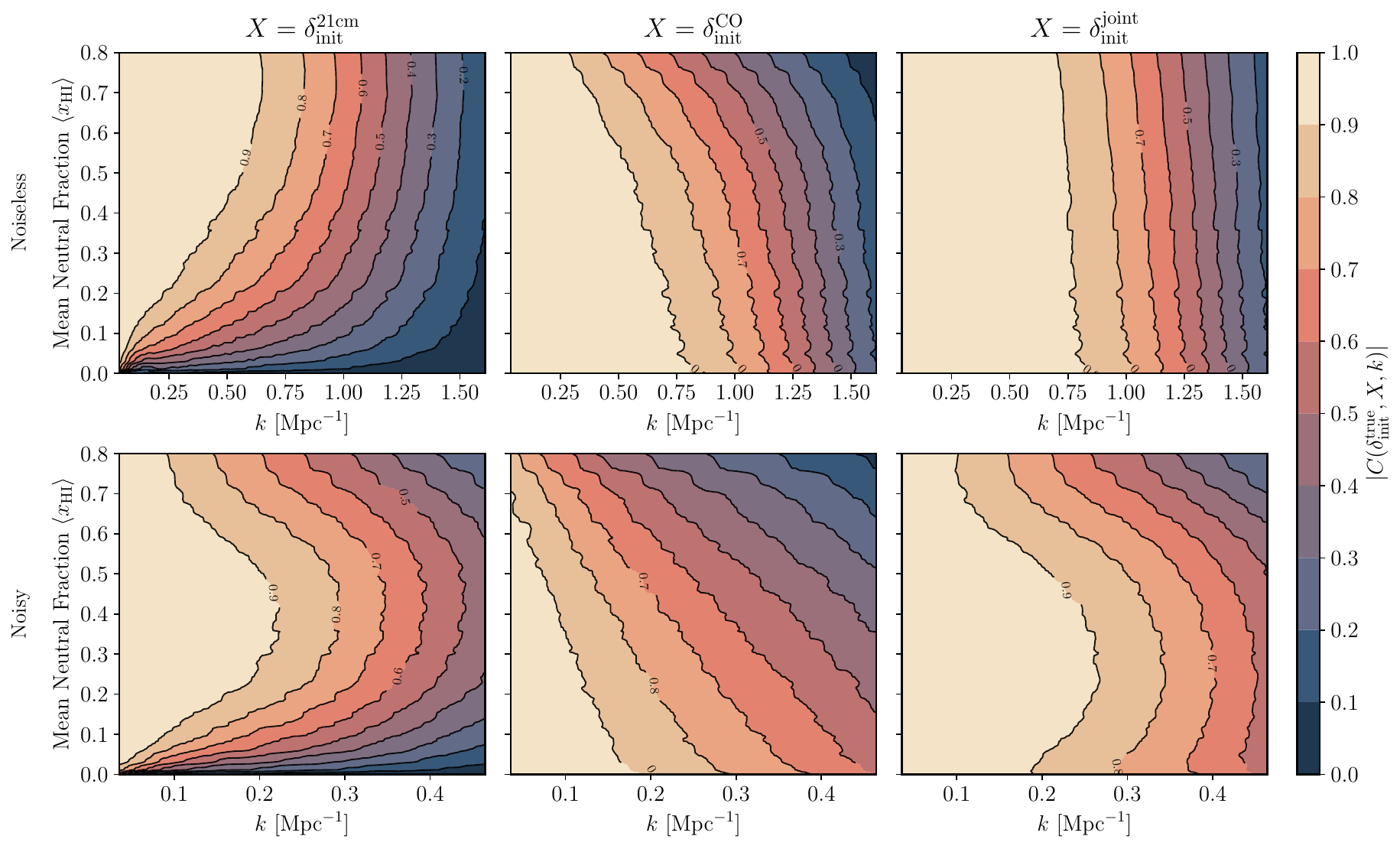}
    \caption{Post-reconstruction cross-correlation coefficient $|C(\dtrue,X;k)|$ between the true and reconstructed initial density fields, shown as a function of wavenumber $k$ and mean neutral fraction $\langle x_{\HI}\rangle$ for 500 independent test realisations. Columns show reconstructions using 21-cm-only (left), CO-only (middle), and joint 21-cm + CO inputs (right). The top row shows the noiseless case, while the bottom row includes the instrumental response and thermal noise adopted for SKA1-Low and COMAP-ERA. Contours indicate levels of constant cross-correlation.}
    \label{fig:cross_comparison_all}
\end{figure*}
In this section, we first benchmark our U-Net by reconstructing the initial density field from the EoR 21-cm and CO line intensity maps, separately and jointly, in both noiseless and noisy settings in Section~\ref{subsec:results_reconstruction}. We then use these reconstructions to derive cosmological parameter constraints in Section~\ref{sec:cosmological_constraints}.

\subsection{Reconstruction of the Initial Density Field}
\label{subsec:results_reconstruction}
To evaluate the performance of our U-Net, we analyse two representative test samples with contrasting neutral fractions set by their underlying cosmological parameters. This setup lets us (i) quantify how reconstruction fidelity varies with the ionisation state of the IGM and (ii) isolate which input carries the dominant information by comparing 21-cm-only, CO-only, and joint-inputs, thereby assessing the incremental gain from combining the tracers. Hereafter, we denote the 21-cm-only, CO-only, and joint-input reconstructions as $\dinitXXI$, $\dinitCO$, and $\dinitJ$, respectively.

\subsubsection{Noiseless Analysis}
Our reconstruction results for the two cases are summarised in Figure~\ref{fig:summary_plot_neutral} (mostly neutral) and Figure~\ref{fig:summary_plot_ionized} (mostly ionised). In each figure, the top row shows representative 2D slices of the input 21-cm ($\dXXI$) and CO ($\dCO$) maps together with the ground-truth initial density ($\dtrue$). The second row shows U-Net reconstructions using 21-cm-only ($\dinitXXI$), CO-only ($\dinitCO$), and joint 21-cm + CO inputs ($\dinitJ$). The third row shows the one-point probability density function (PDF) of the reconstructed initial density field. The bottom row reports scale-dependent statistics: the power spectra of the true and reconstructed initial density fields (left) and the cross-correlations with the ground truth, $|C(\dtrue, X)(k)|$ (right).

\paragraph*{Low Ionisation Fraction:}
When the IGM is largely neutral, as shown in Figure~\ref{fig:summary_plot_neutral}, ionised regions are small and relatively rare, so the 21-cm brightness temperature closely follows the underlying matter density. This is clear from the morphological similarity between $\dXXI$ and $\dtrue$. Consequently, the \XXIonly{} reconstruction recovers most of the structure in the initial density field as evident through a visual comparison between $\dtrue$ and $\dinitXXI$. Quantitatively, the power spectra of the true initial density field $P(k; \dtrue)$ and the reconstructed field $P(k; \dinitXXI)$ show excellent agreement across most scales, with deviations confined to the smallest scales (large $k$-modes). The post-reconstruction cross-correlation remains high, with $|C(\dtrue, \dinitXXI)| \gtrsim 0.65$ for $k \lesssim 1\, {\rm Mpc}^{-1}$, representing a substantial improvement over the pre-reconstruction cross-correlation, $|C(\dtrue, \dXXI)|$ $\gtrsim 0.10$, on the same scales.

By contrast, the CO signal is relatively weak and spatially sparse in this sample, reflecting the small number of luminous galaxies already in place. The \COonly{} reconstruction ($\dinitCO$) therefore underperforms the \XXIonly{} case ($\dinitXXI$). The power spectrum of the reconstructed initial density field $P(k; \dinitCO)$ underestimates the true power spectrum $P(k; \dtrue)$ on intermediate and small scales ($k \gtrsim 0.3\,{\rm Mpc}^{-1}$), and the post-reconstruction cross-correlation $|C(\dtrue, \dinitCO)|$ drops significantly compared to $|C(\dtrue, \dinitXXI)|$. Even so, $|C(\dtrue, \dinitCO)|$ $\gtrsim 0.35$ represents a clear gain over the pre-reconstruction cross-correlation $|C(\dtrue, \dCO)|$ $\gtrsim 0.15$ for $k \lesssim 1\, {\rm Mpc}^{-1}$.

The \Joint{} reconstruction combines the strengths of both tracers and yields the highest overall reconstruction fidelity. The post-reconstruction cross-correlation $|C(\dtrue, \dinitJ)|$ remains $\gtrsim 0.70$ for $k \lesssim 1\,{\rm Mpc}^{-1}$ and exceeds either single-tracer post-reconstruction cross-correlation across all scales probed. Its recovered power spectrum also most closely tracks both the shape and amplitude of the true power spectrum. 

\paragraph*{High Ionisation Fraction:}
When the IGM is largely ionised, as shown in Figure~\ref{fig:summary_plot_ionized}, the situation is markedly different. The 21-cm brightness temperature field no longer uniformly traces the matter density because large ionised bubbles erase the signal in and around overdense regions. As a result, the \XXIonly{} reconstruction recovers structure in regions that are not fully ionised. However, within ionised bubbles, the reconstruction fails to recover substructures and collapses to uniform overdensities. This is consistent with the under-representation of the high-density tail in the one-point PDF of $\dinitXXI$. Quantitatively, the post-reconstruction cross-correlation degrades relative to the mostly neutral sample, with $|C(\dtrue, \dinitXXI)| \gtrsim 0.35$ for $k \lesssim 1\,{\rm Mpc}^{-1}$ (vs.\ $\gtrsim 0.65$ previously in Figure~\ref{fig:summary_plot_neutral}). The power spectrum of the reconstructed initial density field $P(k; \dinitXXI)$ also underestimates the true power spectrum $P(k; \dtrue)$ on intermediate and small scales ($k \gtrsim 0.2\,{\rm Mpc}^{-1}$).

\begin{figure}
    \centering
     \includegraphics[width=\linewidth]{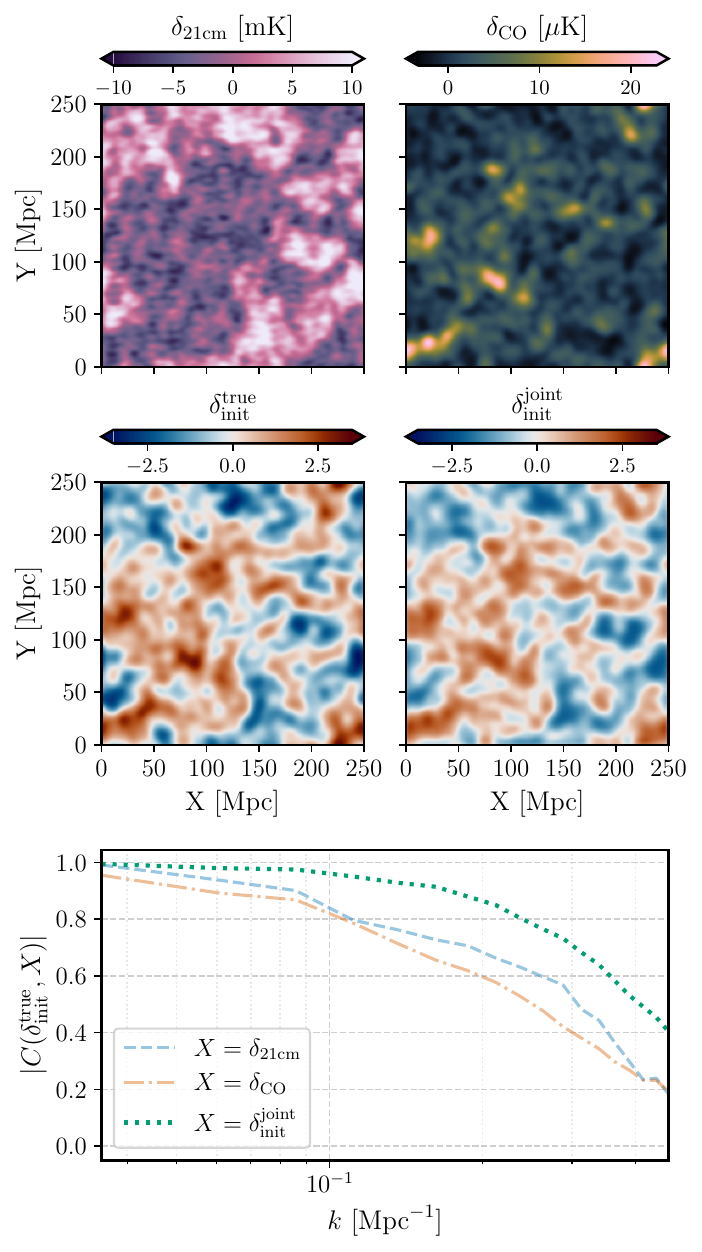}
    \caption{Reconstruction of the initial density field for a noisy sample with $\langle x_{\HI}\rangle \approx 0.12$. The top row shows the input fields: ($\dXXI$ and $\dCO$). The middle row shows the ground truth, $\dtrue$ (left) and the reconstructed initial density field, $\dinitJ$ (right) when joint 21-cm + CO maps are used as input to the U-Net. The bottom row shows the cross-correlation of the true and reconstructed initial density fields $|C (\dtrue , X)|$.}
    \label{fig:noisy_analysis_img}
\end{figure}
In contrast, the CO emission traces the galaxies that drive reionisation and preferentially traces overdense peaks that have ionised their surroundings. The $\dCO$ slice shows strong spatial correlation with $\dtrue$, and the \COonly{} reconstruction achieves substantially higher post-reconstruction cross-correlation compared to the \XXIonly{} case with $|C(\dtrue, \dinitCO)|$ $\gtrsim 0.75$ for $k \lesssim 1\,{\rm Mpc}^{-1}$  vs.\ $|C(\dtrue, \dinitXXI)| \gtrsim 0.35$ on the same scales. Its power spectrum $P(k; \dinitCO)$ also follows the shape and amplitude of $P(k; \dtrue)$ more closely on large and intermediate scales, with differences confined to the highest $k$ modes.

Once again, the \Joint{} reconstruction outperforms both single-tracer inputs. The CO emission helps restore information lost to ionisation in the 21-cm map, while the 21-cm signal fluctuations continue to provide information about underdense regions. 

\paragraph*{Dependence on Ionisation State:}
To quantify how each tracer contributes to reconstruction across ionisation conditions, we apply the trained network to 500 independent test samples spanning a wide range of mean neutral fractions $\langle x_{\HI}\rangle$. Thanks to the fast inference time ($\sim 1$ s per volume), we can map the post-reconstruction coherence $\lvert C(\dtrue, X)\rvert$ as a joint function of wavenumber  $k$ and $\langle x_{\HI}\rangle$, for $X\in\{\dinitXXI,\dinitCO,\dinitJ\}$, as shown in Figure~\ref{fig:cross_comparison_all}. The top row shows the noiseless case, while the bottom row shows the corresponding reconstruction after including the instrumental response and thermal noise described in Section~\ref{sec:inst_noise}.

In the noiseless case, the trends inferred from representative realisations above persist across the full test set. The \XXIonly{} reconstruction performs increasingly well towards a more neutral IGM, where the 21-cm brightness temperature field more closely traces the underlying density field. Its performance progressively degrades towards lower $\langle x_{\HI}\rangle$, as ionised regions erase the 21-cm signal and hence information about the underlying density field. Conversely, the \COonly{} reconstruction performs more strongly towards increasingly ionised conditions, where CO emission traces the overdense, galaxy-rich regions associated with the sources driving reionisation. The joint reconstruction combines these complementary regimes and consequently exhibits substantially weaker dependence on $\langle x_{\HI}\rangle$, while maintaining high coherence over a broader range of scales.

The inclusion of observational effects modifies these trends. Most notably, the \XXIonly{} reconstruction no longer improves monotonically towards the most neutral realisations. Instead, the highest coherence occurs over a broad range of intermediate neutral fractions, with the reconstruction degrading towards both more ionised and more neutral states. Observational effects also reduce the range of scales over which the reconstructed fields remain strongly correlated with the true initial conditions, consistent with the loss of small-scale information from finite instrumental resolution and thermal noise.

Despite this degradation, the complementary behaviour of the two tracers remains evident. The \COonly{} reconstruction retains substantial coherence towards lower neutral fractions, where the \XXIonly{} reconstruction becomes less effective. The \Joint{} reconstruction combines these complementary trends and maintains high coherence over a broad range of ionisation states, although its performance becomes more dependent on $\langle x_{\HI}\rangle$ than in the noiseless case. Thus, the benefit of combining tracers that probe complementary aspects of the EoR persists in the presence of the observational effects considered here.

\subsubsection{Noisy Analysis}
\label{subsec:results_noisy}
We now analyse the performance of reconstruction in the presence of instrumental response and thermal noise as modelled in Section~\ref{sec:inst_noise} (SKA1-Low for 21-cm; COMAP-ERA for CO). We consider a single representative sample from the test dataset, and summarise the results in Figure~\ref{fig:noisy_analysis_img}. The top row shows noisy, resolution-matched input 2D slices of $\dXXI$ and $\dCO$. The middle row compares the true initial density field $\dtrue$ with the reconstructed field from joint inputs $\dinitJ$. The bottom panel displays the cross-correlation $|C(\dtrue, X)|$ as a function of $k$, contrasting pre-reconstruction ($X = \{\dXXI, \dCO\}$) and post-reconstruction ($X = \dinitJ$) correlation. 

Beam smoothing and thermal noise suppress small-scale features in both input maps, as seen in the softened morphology of the top-row panels. This, in turn, limits the recovery of high-$k$ modes. Despite these degradations, the U-Net reconstructs the main morphological features of the initial density field as shown in the middle row. Visually, the reconstructed field $\dinitJ$ retains the large-scale overdense and underdense structures of the true initial density field. 

\begin{figure*}
    \centering
    \includegraphics[width=0.98\linewidth]{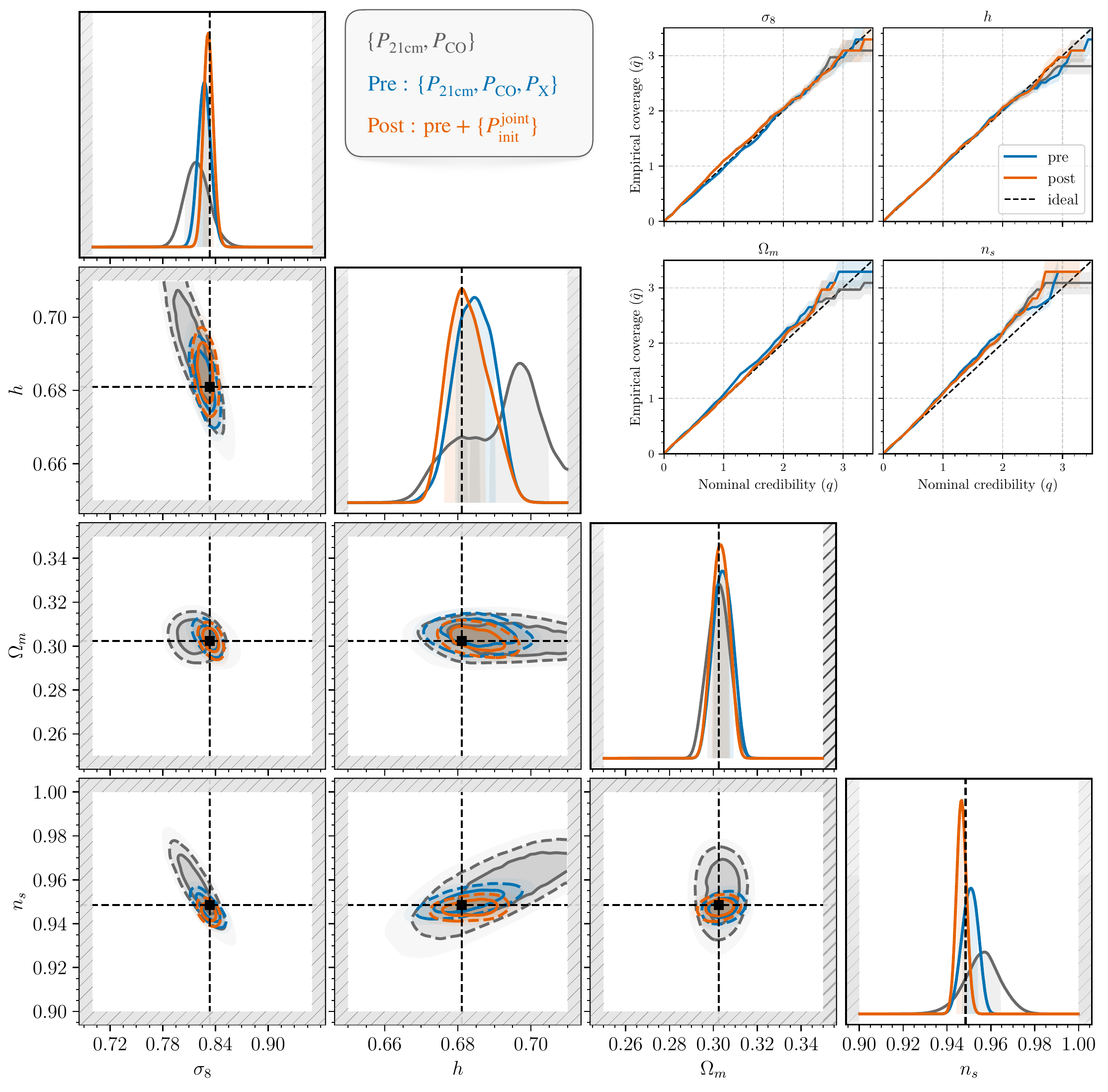}
    \caption{Marginal posterior constraints for a representative noiseless test realisation. Grey, blue, and orange show the auto-power-spectrum-only ($\mathcal{S}_{\rm auto}$), pre-reconstruction ($\mathcal{S}_{\rm pre}$), and post-reconstruction ($\mathcal{S}_{\rm post}$) summaries, respectively. Black dashed lines indicate the true parameter values, and hatched regions denote the prior boundaries. The upper-right panels show the empirical coverage $\hat q$ against the nominal Gaussian-equivalent credibility $q$ for the pre- and post-reconstruction cases; the dashed diagonal denotes ideal calibration.
    }
    \label{fig:param_infer_noiseless}
\end{figure*}

Turning to the cross-correlation statistics, we see that noise suppresses the post-reconstruction cross-correlation relative to the noiseless case. Nevertheless, the bottom panel of Figure~\ref{fig:noisy_analysis_img} shows that the post-reconstruction cross-correlation $|C(\dtrue, \dinitJ)|$ $\gtrsim 0.55$ for $k \lesssim 0.4\,{\rm Mpc}^{-1}$, exceeding pre-reconstruction values $|C(\dtrue, \dXXI)|$ and $|C(\dtrue, \dCO)|$ $\gtrsim 0.25$ at the same scales. Moreover, post-reconstruction $|C(\dtrue, \dinitJ)|$ declines more slowly with $k$ than the pre-reconstruction $|C(\dtrue, \dXXI)|$ and $|C(\dtrue, \dCO)|$, indicating that the network extracts physically meaningful correlations from noise-contaminated data beyond the scales where the raw inputs decorrelate with the initial density field.

Overall, these results show that reconstruction remains effective after incorporating the SKA1-Low and COMAP-ERA instrumental responses and thermal-noise levels considered here, with the highest fidelity retained on large scales. Reconstruction performance degrades towards smaller scales, where finite angular and spectral resolution together with thermal noise limit the available information.

\subsection{Cosmological Parameter Constraints}
\label{sec:cosmological_constraints}
We next turn from reconstruction performance to cosmological inference and quantify the information gained by including the power spectrum of the reconstructed initial density field. Following Section~\ref{sec:summary_sets}, we take $\mathcal{S}_{\rm pre}=\{\pxxi,\pco,\pCross\}$ as our fiducial pre-reconstruction summary and compare it with $\mathcal{S}_{\rm post}=\mathcal{S}_{\rm pre}\cup \{\pinitJ\}$. We additionally retain $\mathcal{S}_{\rm auto}=\{\pxxi,\pco\}$ to quantify the information supplied by the cross-power spectrum itself, and use $\mathcal{S}_{\rm true}$ as the idealised benchmark defined in Section~\ref{sec:summary_sets}.
\begin{figure*}
    \centering
    \includegraphics[width=\linewidth]{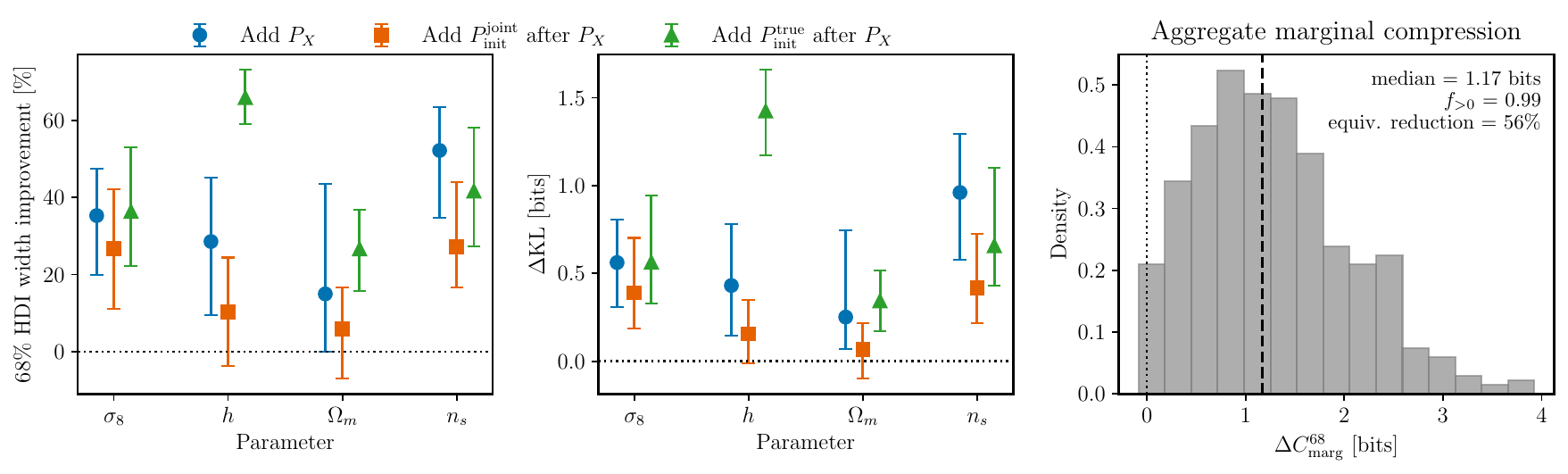}
    \caption{
    Incremental information gain in the noiseless analysis. The left and middle panels show the median improvement in the 68\% HDI width and $\Delta{\rm KL}$, respectively, from adding $P_X$ (blue), adding $P_{\rm init}^{\rm joint}$ after $P_X$ (orange), and adding $P_{\rm init}^{\rm true}$ after $P_X$ (green). Error bars span the 16th--84th percentiles across realisations. The right panel shows the aggregate marginal compression for $\mathcal{S}_{\rm pre}\rightarrow\mathcal{S}_{\rm post}$; the dashed line marks the median and the dotted line zero.
    }
    \label{fig:incremental_info_noiseless}
\end{figure*}
\begin{figure*}
    \includegraphics[width=\linewidth]{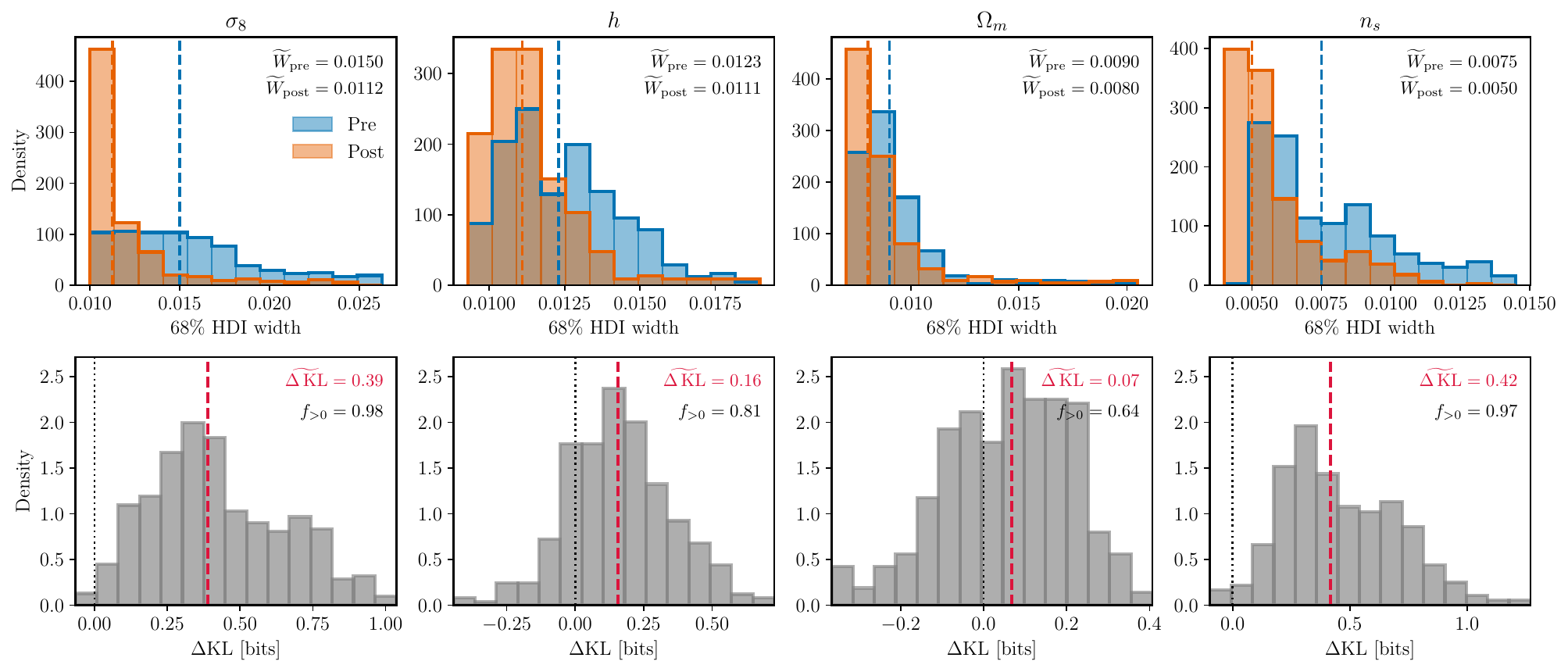}
    \caption{
    Distributions of the marginal posterior diagnostics for the noiseless $\mathcal{S}_{\rm pre}\rightarrow\mathcal{S}_{\rm post}$ comparison. The upper row shows the 68\% HDI widths before (blue) and after (orange) adding $P_{\rm init}^{\rm joint}$, with dashed lines marking the medians. The lower row shows the corresponding $\Delta{\rm KL}$ distributions; the dotted line marks zero and the dashed line the median. For each parameter, the pre- and post-reconstruction widths are evaluated using the same subset of realisations with non-prior-limited 68\% intervals.
    }
    \label{fig:sigma_kl_noiseless}
\end{figure*}

\begin{table}
    \centering
    \caption{Inferred parameters with 16th and 84th percentile credible intervals of the posteriors in Figure~\ref{fig:param_infer_noiseless}, comparing (i) pre- and (ii) post-reconstruction.} 
    \label{Tab:cosmo_noNoise}
    \begin{tabular}{lllll}
     \hline
     Model & $\sigma_8$ & $h$ & $\Omega_{\rm m}$ & $n_{\rm s}$\\
     \hline
     true & $0.833$ & $0.681$ & $0.302$ & $0.948$\\[5pt]
     $\mathcal{S}_{\rm auto}$ & $0.818_{-0.012}^{+0.014}$ & $0.697_{-0.017}^{+0.008}$ & $0.303_{-0.005}^{+0.004}$ & $0.957_{-0.008}^{+0.007}$\\[5pt]
     $\mathcal{S}_{\rm pre}$ & $0.827_{-0.006}^{+0.006}$ & $0.685_{-0.005}^{+0.005}$ & $0.304_{-0.004}^{+0.004}$ & $0.951_{-0.003}^{+0.003}$\\[5pt]
     $\mathcal{S}_{\rm post}$ & $0.832_{-0.005}^{+0.005}$ & $0.681_{-0.004}^{+0.006}$ & $0.303_{-0.003}^{+0.003}$ & $0.947_{-0.002}^{+0.002}$\\
     \hline
    \end{tabular}
\end{table}

\subsubsection{Noiseless Analysis}
\label{sec:noiseless_inference}
We begin with the noiseless setting and perform simulation-based inference using MNRE as described in Section~\ref{sec:sbi_setup}. We train separate estimators for the different power-spectrum summary configurations using 10,000 simulations drawn from the priors in Table~\ref{tab:cosmo_params}, split 80/10/10 into training/validation/test datasets.

\paragraph*{Representative Constraints:}
Figure~\ref{fig:param_infer_noiseless} illustrates the resulting one- and two-dimensional marginal posteriors for a representative test realisation. In addition to the fiducial pre- and post-reconstruction summaries, we show the auto-power-spectrum-only case $\mathcal{S}_{\rm auto}$ to illustrate the effect of including the 21-cm--CO cross-power spectrum $\pCross$. The injected parameter values are indicated by the dashed lines. The inclusion of $P_{\rm X}$ already modifies and, for several parameters, narrows the constraints relative to the auto-power-spectrum-only case. Adding $\pinitJ$ subsequently produces a further contraction of the posterior, most visibly for $\sigma_8$ and $n_{\rm s}$ in this realisation. The inferred parameter values with 16th and 84th percentile uncertainties are listed in Table~\ref{Tab:cosmo_noNoise}.

We next validate our inference using the statistical coverage test introduced in Section~\ref{sec:sbi_setup}. Once the network is trained, we perform inference on 500 samples from the held-out test dataset. Thanks to local amortisation in \texttt{swyft} \citep{https://doi.org/10.5281/zenodo.5043706}, the posterior evaluation for each sample is fast, requiring only $\sim 1\,\mathrm{s}$ per sample. The top-right inset panel of Figure~\ref{fig:param_infer_noiseless} reports empirical coverage versus nominal credibility for all model parameters. The uncertainty interval follows from the finite sample size and is estimated using the Jeffreys interval \citep{Cole_2022}. The black dashed line marks perfect calibration. For both pre- and post-reconstruction summaries, we find an excellent posterior coverage.

\begin{figure*}
    \centering    
    \includegraphics[width=0.98\linewidth]{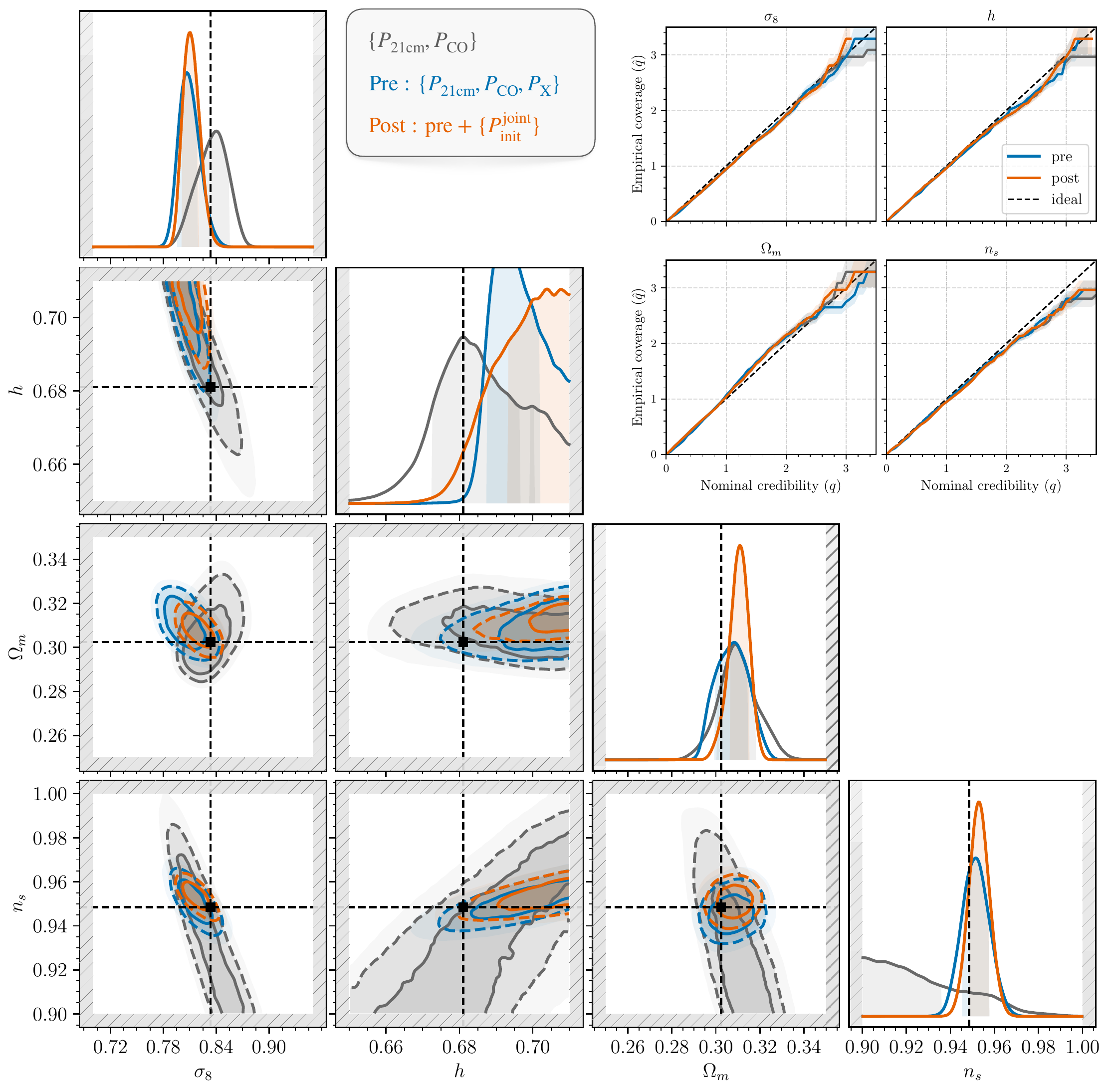}
    \caption{Same as Figure~\ref{fig:param_infer_noiseless}, but including the SKA1-Low and COMAP-ERA instrumental responses and thermal noise.}
    \label{fig:param_infer_noisy}
\end{figure*}

\begin{table}
    \centering
    \caption{Summary of the noiseless $\mathcal{S}_{\rm pre}\rightarrow\mathcal{S}_{\rm post}$ comparison. $\widetilde{W}_{\rm pre}$ and $\widetilde{W}_{\rm post}$ denote the median 68\% HDI widths across realisations. The final column gives the fraction of realisations with $\Delta{\rm KL}>0$.}
    \label{tab:noiseless_summary}
    \begin{tabular}{lccc}
    \hline
    Parameter &
    $\widetilde{W}_{\rm pre}$ &
    $\widetilde{W}_{\rm post}$ &
    $f(\Delta{\rm KL}>0)$ \\
    \hline
    $\sigma_8$ & 0.0150 & 0.0112 & 0.98 \\[3pt]
    $h$        & 0.0123 & 0.0111 & 0.81 \\[3pt]
    $\Omega_{\rm m}$ & 0.0090 & 0.0080 & 0.64 \\[3pt]
    $n_{\rm s}$      & 0.0075 & 0.0050 & 0.97 \\
    \hline
    \end{tabular}
\end{table}

\paragraph*{Ensemble Information Gain:}
We extend this analysis to an ensemble of mock observations to assess whether the improvements in the cosmological parameter constraints persist across realisations. The first two panels of Figure~\ref{fig:incremental_info_noiseless} summarise the change in the 68\% highest-density interval (HDI) widths and the marginal information gain $\Delta{\rm KL}$ for the successive additions of $\pCross$ and $\pinitJ$. We also show the corresponding comparison obtained by replacing $\pinitJ$ with $\pTrue$ as an idealised benchmark. Markers denote the median across realisations, while the error bars span the 16th--84th percentiles and represent realisation-to-realisation scatter.

The cross-power spectrum itself carries substantial additional information: the transition $\mathcal{S}_{\rm auto}\rightarrow\mathcal{S}_{\rm pre}$ produces positive median reductions in the 68\% HDI width and positive median $\Delta{\rm KL}$ for all four parameters considered. For the subsequent transition $\mathcal{S}_{\rm pre}\rightarrow\mathcal{S}_{\rm post}$, the median 68\% HDI widths decrease by approximately $27\%$, $10\%$, $6\%$, and $27\%$ for $\sigma_8$, $h$, $\Omega_{\rm m}$, and $n_{\rm s}$, respectively. The corresponding median information gains are approximately $0.39$, $0.16$, $0.07$, and $0.42$ bits. The improvement is therefore strongest for $\sigma_8$ and $n_{\rm s}$, while $h$ and particularly $\Omega_{\rm m}$ show more modest gains.

The idealised $\pTrue$ benchmark provides a useful reference for assessing how closely the reconstructed power spectrum approaches the constraining power of the true initial-density power spectrum. For all four parameters, adding $\pTrue$ yields larger median improvements than adding $\pinitJ$. The difference is relatively small for $\sigma_8$ and $n_{\rm s}$, whereas it is substantially  larger for $h$ and $\Omega_{\rm m}$, particularly for $h$. Given that the reconstruction fidelity decreases towards smaller scales, the larger gaps for $h$ and $\Omega_{\rm m}$, suggest that their constraints may be more sensitive to small-scale information, which is not as well recovered by the reconstruction.

The right panel of Figure~\ref{fig:incremental_info_noiseless} combines these parameter-by-parameter changes using the factorised marginal-compression statistic defined in Section~\ref{sec:information_gain}. For the addition of $\pinitJ$ after $\pCross$, the distribution has a median $\Delta C_{\rm marg}^{68}=1.17$ bits, with positive compression in $99\%$ of the test realisations. This corresponds to a median reduction of approximately $56\%$ in the factorised product of prior-normalised 68\% marginal widths. As emphasised in Section~\ref{sec:information_gain}, this quantity is a marginal compression diagnostic rather than the volume of the full joint posterior, since correlations among cosmological parameters are not included in its construction.

Figure~\ref{fig:sigma_kl_noiseless} shows the full realisation-to-realisation distributions for the primary comparison $\mathcal{S}_{\rm pre}\rightarrow\mathcal{S}_{\rm post}$, with the corresponding numerical summary given in Table~\ref{tab:noiseless_summary}. To avoid interpreting prior-limited credible intervals as meaningful width improvements, the pre- and post-reconstruction width distributions for each parameter are evaluated using the same subset of realisations for which the 68\% HDIs are not prior limited. The distributions show a systematic shift towards narrower post-reconstruction constraints, with the strongest changes for $\sigma_8$ and $n_{\rm s}$. The $\Delta{\rm KL}$ distributions are also predominantly positive, although the degree of realisation-to-realisation consistency varies between parameters and is weakest for $\Omega_{\rm m}$.

Taken together, these results show that the 21-cm--CO cross-power spectrum provides substantial cosmological information beyond the individual tracer auto-power spectra, while the subsequent addition of $\pinitJ$ provides further, complementary information beyond these pre-reconstruction summaries.

\subsubsection{Noisy Analysis}
\label{sec:noisy_inference}
We next repeat the inference in the noisy setting, incorporating the SKA1-Low and COMAP-ERA observational effects. We use the same cosmological priors and hierarchy of power-spectrum summary sets as in the noiseless analysis, with the summary statistics now constructed from the observationally degraded tracer maps and the corresponding reconstructed initial density field.

\paragraph*{Representative Constraints:}
Figure~\ref{fig:param_infer_noisy} shows the resulting one- and two-dimensional marginal posteriors for a representative noisy test realisation. As in the noiseless case, we compare the auto-power-spectrum-only summary $\mathcal{S}_{\rm auto}$, the fiducial pre-reconstruction summary $\mathcal{S}_{\rm pre}$, and the post-reconstruction summary $\mathcal{S}_{\rm post}$. The inclusion of $\pCross$ modifies and, for several parameters, narrows the constraints relative to the auto-power-spectrum-only case, while adding $\pinitJ$ produces a further contraction of the posterior. We again validate the inference using the statistical coverage test. The empirical versus nominal credibility shown in the upper-right panels of Figure~\ref{fig:param_infer_noisy} remains broadly consistent with the ideal calibration relation for both the pre- and post-reconstruction summaries.

\begin{figure*}
    \centering
    \includegraphics[width=\linewidth]{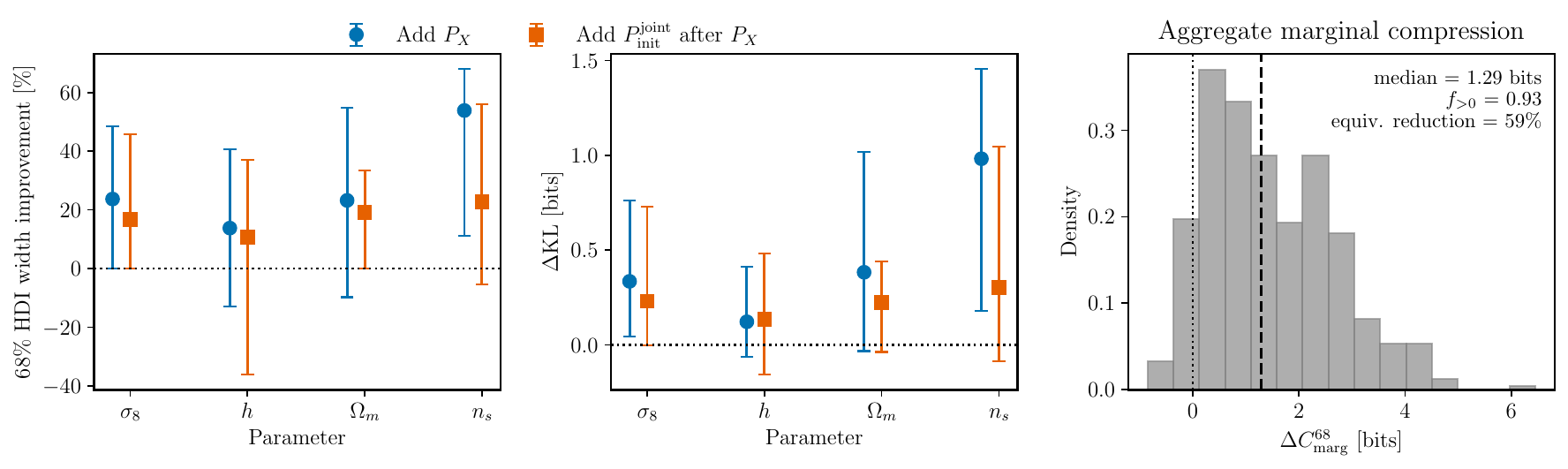}
    \caption{
    Incremental information gain in the noisy analysis. The left and middle panels show the median improvement in the 68\% HDI width and $\Delta{\rm KL}$, respectively, from adding $P_X$ (blue) and adding $P_{\rm init}^{\rm joint}$ after $P_X$ (orange). Error bars span the 16th--84th percentiles across realisations. The right panel shows the aggregate marginal compression for $\mathcal{S}_{\rm pre}\rightarrow\mathcal{S}_{\rm post}$.
    }
    \label{fig:incremental_information_noisy}
\end{figure*}

\paragraph*{Ensemble Information Gain:}
We next extend the analysis to the ensemble of 500 test realisations. The first two panels of Figure~\ref{fig:incremental_information_noisy} summarise the changes in the 68\% HDI widths and marginal information gain $\Delta{\rm KL}$ for the successive additions of $\pCross$ and $\pinitJ$. Markers denote the median across realisations, while the error bars span the 16th--84th percentiles and represent realisation-to-realisation scatter. The cross-power spectrum continues to provide substantial additional information beyond the individual tracer auto-power spectra. For the subsequent transition $\mathcal{S}_{\rm pre}\rightarrow\mathcal{S}_{\rm post}$, the median 68\% HDI widths are reduced for all four parameters considered. The corresponding median information gains are approximately $0.23$, $0.13$, $0.22$, and $0.30$ bits for $\sigma_8$, $h$, $\Omega_m$, and $n_s$, respectively, showing positive ensemble-level gains after the addition of $\pinitJ$.

The right panel of Figure~\ref{fig:incremental_information_noisy} combines these parameter-by-parameter changes using the factorised marginal-compression statistic. For the addition of $\pinitJ$ after $\pCross$, the distribution has a median $\Delta C_{\rm marg}^{68}=1.29$ bits, with positive compression in $93\%$ of the test realisations. This corresponds to a median reduction of approximately $59\%$ in the factorised product of prior-normalised 68\% marginal widths. The distribution exhibits greater realisation-to-realisation scatter than in the noiseless analysis, but remains preferentially shifted towards positive compression.

\begin{figure*}
    \centering
    \includegraphics[width=\linewidth]{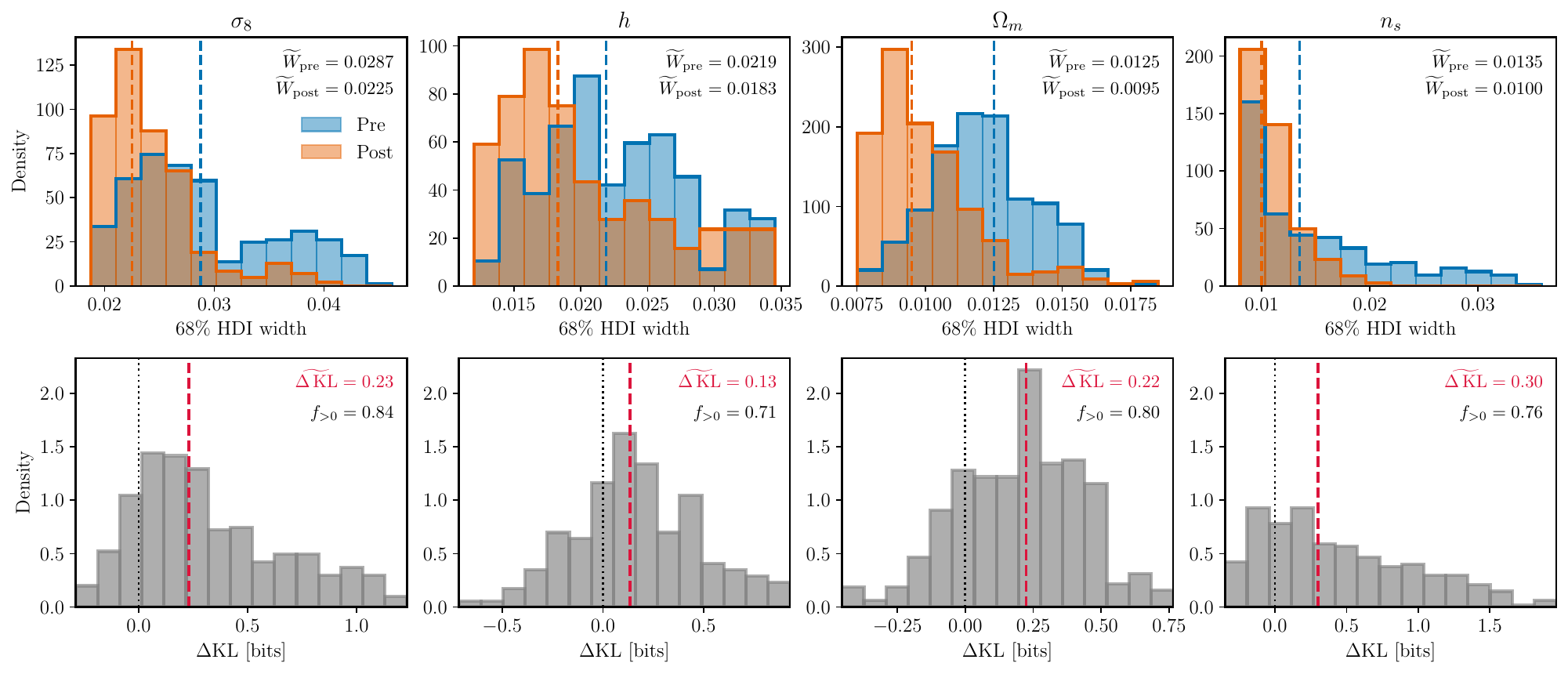}
    \caption{
    Distributions of the marginal posterior diagnostics for the noisy $\mathcal{S}_{\rm pre}\rightarrow\mathcal{S}_{\rm post}$ comparison. The upper row shows the 68\% HDI widths before (blue) and after (orange) adding $P_{\rm init}^{\rm joint}$. The lower row shows the corresponding $\Delta{\rm KL}$ distributions. Dashed lines mark the medians and dotted lines indicate $\Delta{\rm KL}=0$.
    }
    \label{fig:pre_post_noisy}
\end{figure*}

Figure~\ref{fig:pre_post_noisy} shows the full realisation-to-realisation distributions for the primary comparison $\mathcal{S}_{\rm pre}\rightarrow\mathcal{S}_{\rm post}$, with the corresponding numerical summary given in Table~\ref{tab:noisy_summary}. The $\Delta{\rm KL}$ distributions are predominantly positive, although they extend below zero for individual realisations. The degree of realisation-to-realisation consistency varies between parameters, with positive information gain occurring most frequently for $\sigma_8$ and least frequently for $h$.

In the presence of observational effects, the 21-cm--CO cross-power spectrum continues to provide substantial cosmological information beyond the individual tracer auto-power spectra, while the subsequent addition of $\pinitJ$ provides further, complementary information beyond these pre-reconstruction summaries.

\begin{table}
  \centering
  \caption{Inferred parameters with 16th and 84th percentile credible intervals of the posteriors in Figure~\ref{fig:param_infer_noisy}, comparing (i) pre- and (ii) post-reconstruction.}
  \label{Tab:cosmo_Noise}
  \begin{tabular}{lcccc}
    \hline
    Model & $\sigma_8$ & $h$ & $\Omega_{\rm m}$ & $n_{\rm s}$\\
    \hline
    true & $0.833$ & $0.681$ & $0.302$ & $0.948$\\[5pt]
    $\mathcal{S}_{\rm auto}$  & $0.841_{-0.020}^{+0.014}$ & $0.681_{-0.008}^{+0.020}$ &
           $0.309_{-0.008}^{+0.009}$ & $0.900^{+0.035}\,^{\dagger}$\\[5pt]
    $\mathcal{S}_{\rm pre}$ & $0.807_{-0.010}^{+0.011}$ & $0.692_{-0.005}^{+0.009}$ &
           $0.308_{-0.008}^{+0.006}$ & $0.952_{-0.006}^{+0.005}$\\[5pt]
    $\mathcal{S}_{\rm post}$ & $0.811_{-0.009}^{+0.009}$ & $0.704_{-0.010}^{+0.006}$ &
           $0.311_{-0.004}^{+0.004}$ & $0.953_{-0.004}^{+0.004}$\\
    \hline
  \end{tabular}
  \vspace{2pt}
  \parbox{\columnwidth}{\footnotesize \textit{Notes.} $^{\dagger}$ One-sided 68\% credible interval i.e.\ the lower error is prior-limited at the prior bound.}
\end{table}


\section{Summary}
\label{sec:summary}
This work demonstrates that tomographic maps of the Epoch of Reionisation (EoR) can be used to reconstruct the initial density field and, in turn, improve cosmological parameter constraints. We train a three-dimensional residual U-Net to recover the initial density field from paired 21-cm and CO(1--0) line-intensity cubes at $z=8$. The network operates on $128^3$ grids with two input channels (21-cm and CO) and produces a single-channel initial density field reconstruction. 

We first assess reconstruction fidelity across ionisation conditions set by the underlying cosmology. Using noiseless line-intensity maps, the \XXIonly\ reconstructions are highly correlated with the true initial density field when the IGM is mostly neutral, while \COonly\ reconstructions perform better when the IGM is highly ionised. Combining the tracers yields consistently superior performance: the post-reconstruction coherence $|C(\dtrue, \dinitJ)(k)|$ remains $\gtrsim 0.75$ for $k\lesssim 1\,{\rm Mpc}^{-1}$ across a wide range of ionisation states. 

Extending this comparison across 500 independent realisations confirms the complementary dependence of the two tracers on the ionisation state. Although observational effects weaken this dependence, particularly for the 21-cm-only reconstruction, the joint reconstruction remains comparatively robust across ionisation states.
\begin{table}
    \centering
    \caption{
    Summary of the noisy $\mathcal{S}_{\rm pre}\rightarrow\mathcal{S}_{\rm post}$ comparison. $\widetilde{W}_{\rm pre}$ and $\widetilde{W}_{\rm post}$ denote the median 68\% HDI widths across realisations. The final column gives the fraction of realisations with $\Delta{\rm KL}>0$.
    }
    \label{tab:noisy_summary}
    \begin{tabular}{lccc}
    \hline
    Parameter &
    $\widetilde{W}_{\rm pre}$ &
    $\widetilde{W}_{\rm post}$ &
    $f(\Delta{\rm KL}>0)$ \\
    \hline
    $\sigma_8$ & 0.0287 & 0.0225 & 0.84 \\[3pt]
    $h$        & 0.0219 & 0.0183 & 0.71 \\[3pt]
    $\Omega_{\rm m}$ & 0.0125 & 0.0095 & 0.80 \\[3pt]
    $n_{\rm s}$      & 0.0135 & 0.0100 & 0.76 \\
    \hline
    \end{tabular}
\end{table}

We then translate these reconstruction gains into cosmological information using simulation-based inference with marginal neural ratio estimation (MNRE). Starting from the individual 21-cm and CO auto-power spectra, we assess the successive information gained from the 21-cm--CO cross-power spectrum and the reconstructed initial-density power spectrum. Across 500 noiseless test realisations, the cross-power spectrum provides substantial additional information, while the subsequent addition of $P_{\rm init}^{\rm joint}$ reduces the median 68\% HDI widths by approximately $27\%$, $10\%$, $6\%$, and $27\%$ for $\sigma_8$, $h$, $\Omega_m$, and $n_s$, respectively. The corresponding marginal $\Delta{\rm KL}$ distributions are predominantly positive, while the aggregate marginal-compression statistic has a median of $\Delta C_{\rm marg}^{68}=1.17$ bits, corresponding to a $56\%$ reduction in the factorised product of prior-normalised marginal widths. Posterior-coverage tests remain broadly consistent with the nominal credibility levels.

We finally incorporate finite angular and spectral resolution and thermal noise appropriate to the SKA1-Low and COMAP-ERA observational assumptions adopted here. As expected, these effects broaden the absolute parameter constraints and suppress small-scale reconstruction coherence. Nevertheless, adding $P_{\rm init}^{\rm joint}$ after the tracer cross-power spectrum continues to improve the ensemble constraints, with narrower median 68\% HDI widths and positive median $\Delta{\rm KL}$ for all four parameters considered. The aggregate marginal compression has a median $\Delta C_{\rm marg}^{68}=1.29$ bits and is positive for $93\%$ of the test realisations, corresponding to a $59\%$ reduction in the factorised marginal-volume proxy. Thus, within the observational assumptions considered here, the reconstruction retains cosmological information beyond the selected tracer auto- and cross-power spectra.

Conceptually, our results illustrate that cosmological information that is straightforwardly encoded in the initial density field becomes redistributed through non-linear gravitational evolution and astrophysical processes in EoR line-intensity observables. By reconstructing the initial field with a multi-channel U-Net and then summarising it at the power-spectrum level, we effectively ``linearise'' part of this information, leading to improved cosmological constraints.

This work can be extended in several directions to strengthen the approach. First, we have held EoR astrophysics fixed; allowing key astrophysical parameters to vary and marginalising over them would test robustness to galaxy and ionisation modelling and may reveal new degeneracies with $\{\sigma_8, h, \Omega_{\rm m}, n_{\rm s}\}$. Second, moving from a single redshift to multi-redshift training should exploit tomographic evolution, improving reconstructions by leveraging temporal correlations. Third, incorporating realistic observational effects such as continuum foregrounds, line interlopers, and $k_{\parallel}$ filtering will be essential for survey realism. Finally, we have focused on point estimates of the initial density field; however, extending to uncertainty-aware, field-level posteriors would propagate reconstruction uncertainty more completely into cosmological inference \citep{savchenko2024meanfieldsimulationbasedinferencecosmological}. We leave these developments to future work, with the goal of making the reconstruction framework more robust and closer to survey-ready applications.

\section*{Acknowledgements}
We thank the Center for Information Technology of the University of Groningen for their support and for providing access to the Hábrók high performance computing cluster. GS was supported by a CIERA Postdoctoral Fellowship, with additional support provided by NSF through grants AST-2108230 and AST-2307327; by NASA through grants 21-ATP21-0036 and 23-ATP23-0008; and by STScI through grant JWST-AR-03252.001-A. 

\section*{Data Availability}
The deep-learning framework developed in this work, \texttt{CosmicRewind}, is publicly available at \url{https://github.com/anchal-009/CosmicRewind}. The data underlying this article will be shared on reasonable request to the corresponding author.



\bibliographystyle{mnras}
\bibliography{references} 




\appendix


\bsp	
\label{lastpage}
\end{document}